Characterizing the Galactic and Extragalactic Background Near Exoplanet Direct Imaging Targets

Authors: M. Cracraft[1], R. J. De Rosa[2], W. Sparks[3], V. P. Bailey[4], M. Turnbull[5]

Abstract: As more missions attempt to directly image and characterize exoplanets orbiting nearby sun-like stars, advance characterization of possible contaminating background sources becomes more important and can impact target selection. This paper describes an exploration of the Hubble Source catalog, Gaia catalog and Besançon galaxy simulations in order to determine the likelihood of having a contaminating source in the background of a set of high proper motion stars in the expected timeframe of observations for the Nancy Grace Roman Space Telescope Coronagraphic Instrument. The analysis shows that for most of the targets, there is a very low possibility of a star falling within the CGI field of view, but that at low galactic latitudes where there is a greater density of sources, faint stellar background sources could be a concern.

1. Introduction

There is a growing impetus for spaceborne missions to directly image and characterize Earth-sized and larger exoplanets orbiting nearby sunlike stars. Advance characterization of confounding background sources in near-future planet hunting fields has the potential to critically impact target selection and observing strategy of these missions. Here, we describe such a project motivated by potential observing programs to use the Nancy Grace Roman Space Telescope (Roman) Coronagraphic Instrument (CGI), but with broad application to other telescopes and missions.

Since astrophysical background sources may be confused with exoplanets orbiting target stars, and hence hamper our ability to identify and characterize such exoplanets, we sought to determine both empirically and statistically the likelihood of encountering background "contaminating" sources. To do this, we extracted images from the Hubble Space Telescope (HST) archives of regions close to target stars which had been observed by HST in order to empirically check for any nearby potentially contaminating stars, galaxies or other astrophysical sources near the target stars, projected forward to the era of the Roman observations. We also examine the Gaia DR2 catalog to determine (a) whether any identified Gaia stars would be near enough to interfere with Roman CGI observations and (b) compare the Gaia catalog to a set of Besançon galaxy simulated catalogs to determine how well that statistical model reproduced the empirical Gaia catalogs at the explicit pointings represented by the CGI targets, and hence tie the empirical likelihood of contamination to a more widely applicable statistical approach.

CGI will observe both exoplanets and circumstellar disks during its technology demonstration phase. As the highest priority circumstellar disk targets (eg: HR 4796), and self-luminous exoplanet targets (eg: HR 8799), already have deep adaptive optics imaging that can be used to identify background contaminants, we concentrate here instead on the potential reflected-light exoplanet targets (Table 1). These targets all host known exoplanets, which were detected by radial velocity (RV) but have not yet been directly imaged.

1) Space Telescope Science Institute, 3700 San Martin Drive, Baltimore MD 21218
2) European Southern Observatory, Alonso de Córdova 3107, Vitacura, Santiago, Chile
3) Space Telescope Science Institute (Emeritus) and SETI Institute
4) Jet Propulsion Laboratory, California Institute of Technology, 4800 Oak Grove Dr, Pasadena CA, 91109
5) SETI Institute 189 Bernardo Ave, Suite 200. Mountain View, CA 94043, United States

*Table 1 Properties of potential CGI targets: J2000 coordinates in RA and Dec, Galactic Latitude and Longitude and proper motions; Parallax, Proper Motion in RA and Dec (mas/yr). These systems host known radial velocity- detected planets, some of which might be detectable with CGI. All coordinates listed were taken from SIMBAD. The last column indicates whether we have Hubble data for each target.*

| Target | RA (h:m:s) | Dec (d:m:s) | Gal. Lat (deg) | Gal. Long. (deg) | Parallax (mas) | PM RA (mas/yr) | PM Dec (mas/yr) | HST |
|---|---|---|---|---|---|---|---|---|
| 14 Her | 16 10 24.315 | +43 49 03.499 | 46.94 | 69.17 | 55.74 | 132.02 | -296.46 | |
| 47 Uma | 10 59 27.974 | +40 25 48.922 | 63.37 | 175.78 | 72.45 | -317.64 | 55.01 | |
| 55 Cnc | 08 52 35.811 | +28 19 50.957 | 37.70 | 196.79 | 79.43 | -485.87 | -233.65 | yes |
| ε Eri | 03 32 55.845 | -09 27 29.731 | -48.05 | 195.84 | 310.94 | -975.17 | 19.49 | yes |
| HD 142 | 00 06 19.175 | -49 04 30.682 | -66.39 | 321.59 | 38.89 | 575.29 | -39.19 | |
| HD 39091 | 05 37 09.885 | -80 28 08.831 | -29.78 | 292.51 | 54.71 | 311.19 | 1048.85 | |
| HD 114613 | 13 12 03.183 | -37 48 10.879 | 24.89 | 307.42 | 49.27 | -381.63 | 46.09 | |
| HD 134987 | 15 13 28.667 | -25 18 33.655 | 27.39 | 339.19 | 38.17 | -400.31 | -75.16 | |
| HD 154345 | 17 02 36.404 | +47 04 54.763 | 37.66 | 73.02 | 54.66 | 123.20 | 853.74 | |
| HD 160691 | 17 44 08.704 | -51 50 02.591 | -11.49 | 340.06 | 64.08 | -15.31 | 190.92 | |
| HD 190360 | 20 03 37.405 | +29 53 48.495 | -0.67 | 67.41 | 62.44 | 683.34 | -525.68 | |
| HD 192310 | 20 15 17.392 | -27 01 58.714 | -29.40 | 15.62 | 113.65 | 1242.53 | -181.04 | |
| HD 217107 | 22 58 15.541 | -02 23 43.387 | -53.32 | 70.47 | 49.82 | -7.11 | -14.78 | |
| HD 219134 | 23 13 16.975 | +57 10 06.0765 | -3.20 | 109.90 | 153.08 | 2074.52 | 294.94 | |
| τ Ceti | 01 44 04.083 | -15 56 14.927 | -73.44 | 173.10 | 273.96 | -1721.05 | 854.16 | yes |
| υ And | 01 36 47.842 | +41 24 19.644 | -20.67 | 132.00 | 74.12 | -173.33 | -381.8 | yes |

2. Observational and Statistical Approach

In order to determine the likelihood of background contamination of CGI observations empirically, a variety of catalogs and archived images were used to examine the background regions of the selected targets.

We checked whether there is any background source within the field of view of the target star at the time when the observations would likely be taken. For this document, the year 2026 is the assumed launch date and fall of 2026 is the likely date of observation. There are several regions to be considered for the CGI instrument.

The high-contrast region for the primary observing mode (Band 1 with the Hybrid Lyot Coronagraph) has a diameter of 0.9 arcseconds (575 nm with outer working angle $\lambda/D = 9$). The high-contrast region for the secondary imaging mode Band 4 with the Shaped Pupil "Wide" FOV Coronagraph has diameter 2.9 arcseconds (imaging at 825 nm with $\lambda/D = 20$). The flux ratio detection limit in these observing modes is $10^{-8} - 10^{-9}$. The non-coronagraphic field of view has diameter 7.2 arcseconds, but high-contrast observations cannot be taken in non-coronagraphic mode. Hence, the coronagraphic field of view is the most important region to be clear of background sources as well as 'glints' from nearby objects. No stars of equal brightness to the target star may be present within about 45" of the target, with increasingly stringent limits at smaller separations: nothing ~5-7 magnitudes fainter within 10" of the region of interest, nothing ~15 magnitudes fainter within 2" of the region of interest.

Initially we examined the Hubble Source Catalog (HSC) and Hubble Legacy Archive (HLA) to see if there were images available covering the positions of the targets in current times as well as dates in 2026, taking into account the expected proper motion of the target stars. Since there was only Hubble data currently existing for a few targets, as listed in Table 1, the Gaia DR2 catalog was examined for the full target list. The Gaia catalogs provide empirical relatively uniform statistics on the number and density of sources near the targets, whereas even where Hubble data exists, it is heterogeneous. The Hubble Source Catalog is derived from all existing observations, from a variety of observing programs, utilizing different instruments, filters and exposure times, making for an inhomogeneous data set. By combining information from both Gaia and HST (where available), a better understanding of the expected characteristics of the background of the target stars and nearby regions can be obtained. The Besançon statistical model for Galaxy star count simulations was also examined, to compare the predicted star counts at different galactic latitudes to the observed values found with Gaia. The comparison between Gaia and Besançon allows us to use the Besançon simulations to extrapolate into fainter magnitudes than are found in the Gaia catalogs. The section below describes an in-depth exploration of the region of sky near Upsilon Andromeda (υ And), one of the few sources that was well covered in both the HSC and Gaia, to illustrate our methods.

3. A case study: exploration of Upsilon Andromeda

This exploration started by examining the Mikulski Archive for Space Telescopes (MAST) contents for Hubble data on the target. By calculating the position of υ And in 2026 taking into account the proper motion, it can be determined whether there is usable data covering the 2026 position and hence reveling the (future) background of the target star. The image below, Figure 1, shows a MAST image (DSS background) with a red cross-hair in the middle of the star image, with coordinates that match the J2000 coordinates on January 1, 2000, and second red cross-hair to the lower right marking the January 1, 2026 coordinates. The search for Hubble data for direct background observations should focus on the position calculated in 2026. As a double-check on the calculated 2026 position, the Gaia data was overplotted on the MAST DSS image. Since the Gaia DR2 data have a timeframe of approximately July of 2015, it should show consistent motion between the 2000 and 2026 dates. The pink square in the middle is a Gaia point source Gaia DR2 348020448377061376, which maps to υ And in the Gaia timeframe. This does show consistent motion and is a good check on the calculations of the 2026 position. While there is a STIS image centered on υ And taken in 2011, the future position (2026) is beyond the edge of any image currently existing in the Hubble data sets.

Table 2 *Positions of Upsilon Andromeda in 2000 and 2026. The J2000 position and proper motion were found in SIMBAD and calculated out to the year 2026 in J2000 coordinates.*

| Coordinate | 2000 Position | Proper Motion (mas/yr) | 2026 position |
|---|---|---|---|
| RA | 1:36:47.842 | -173.33 | 1:36:47.442 |
| Dec | +41:24:19.644 | -381.8 | +41:24:09.72 |

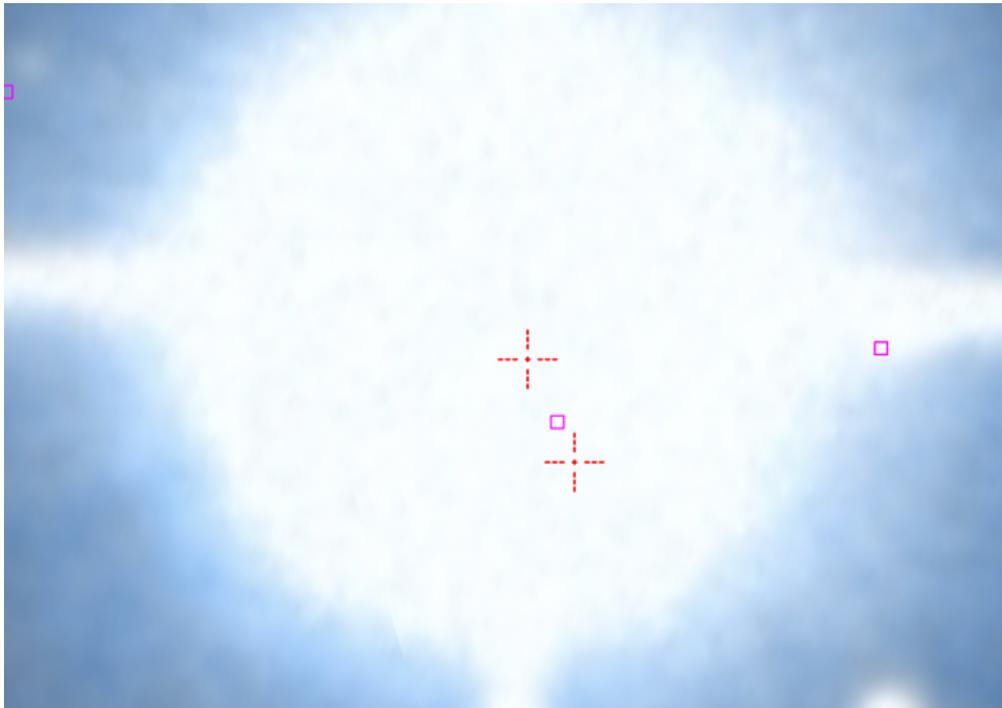

Figure 1 *Current and future positions of υ And on a DSS background. The red cross in the middle of the image marks the 2000 (January 1) coordinates, while the red cross to the lower right mark the 2026 coordinates. The pink box in between marks the Gaia position from July 2015. This shows the movement of the star over 26 years. The star to the right (pink box to the far right) is a potential contaminating star in the Gaia catalog*

Where Hubble data exist, the MAST archive shows the footprint of all observations near or on the target, and allows a user to narrow down which data to examine by instrument and filter amongst other parameters. The approach taken here was to find several images in the field of interest that had consistent cameras and filters that gave large enough regions to allow for some statistical analysis and comparisons to the Gaia data set. For υ And, there exists a set of three separated regions taken with the ACS camera, F775W filter that have a catalog of stars in the HSC.

These image footprints are shown in Figure 2. The footprints outlined in orange show where images are available in the HLA, while the red dots indicate the identified ACS/F775W HSC sources. The circle around the central target has a radius of 18 arcminutes. These three fields of view had processed images in the HLA. These fields combined 15 exposures, 31 exposures, and 30 exposures, for total exposure times of 9391, 16797 and 16302 seconds respectively. Combined images of this depth have very few remaining image artifacts such as cosmic rays. With all three images combined the resulting catalog extracted from the HSC contains over 2300 sources. The catalog can be filtered to include only those sources identified in multiple images in order to exclude possible remaining spurious sources. This process results in approximately 50 sources brighter than 25th magnitude in F775W detected per square arcminute in these combined regions.

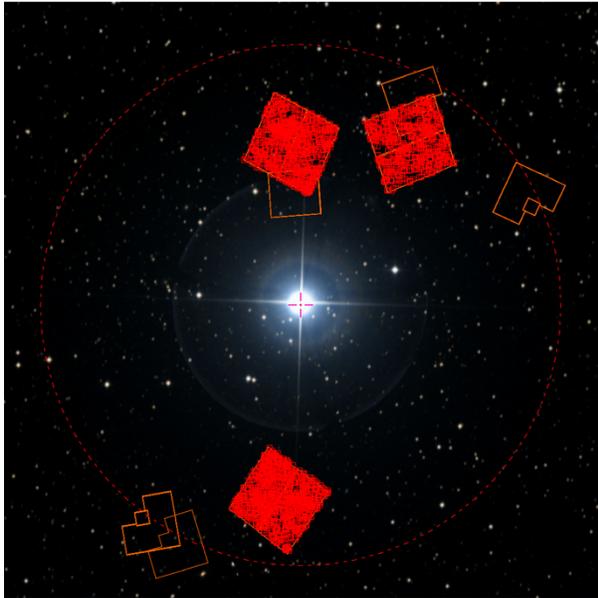

Figure 2 Footprint of υ And Hubble data in the HLA, with the ACS/F775W HSC sources shown in red. The open wedge shapes are the outline of WFPC2 images. The circle to look for sources has an 18 arcminute radius.

Using the data found in the Hubble source catalog, we can plot a histogram of the sources found in the ACS F775W image set.

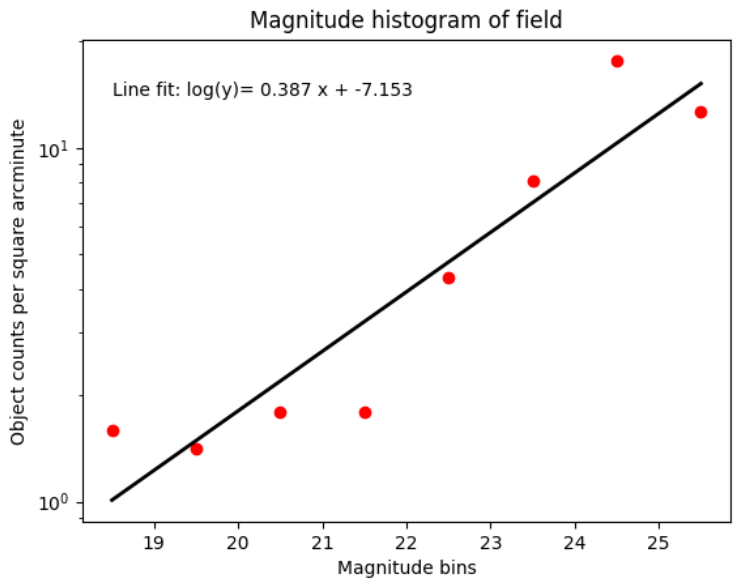

Figure 3 Histogram of the ACS F77W data from the Hubble source catalog. Y axis is number of sources found per square arcminute.

This is useful information from Hubble data, but as can be seen in Table 1, very few of the targets chosen actually have such extensive Hubble data. The Gaia DR2 offers a complementary catalog to the HSC, albeit to a depth much less than typical HST imaging. Gaia is complete to approximately 20 magnitudes while HST finds sources to magnitude 24 - 25. The advantage of Gaia DR2 is that it provides a consistent, relatively uniform, empirical dataset for all targets. All of the data was taken in the same manner and with the same filters. In order to determine stellar densities near the targets and thus the likelihood of having a contaminating source in a given observation, the Gaia DR2 release was therefore used to examine the background area of the targets in a 0.3 degree radius area centered on a target star.

With the Gaia data, color magnitude diagrams can be created for those regions using the different Gaia passbands, G (green), GBP (blue) and GRP (red) as shown here: https://www.cosmos.esa.int/web/gaia/iow_20180316. An example of this color magnitude diagram is shown in Figure 4.

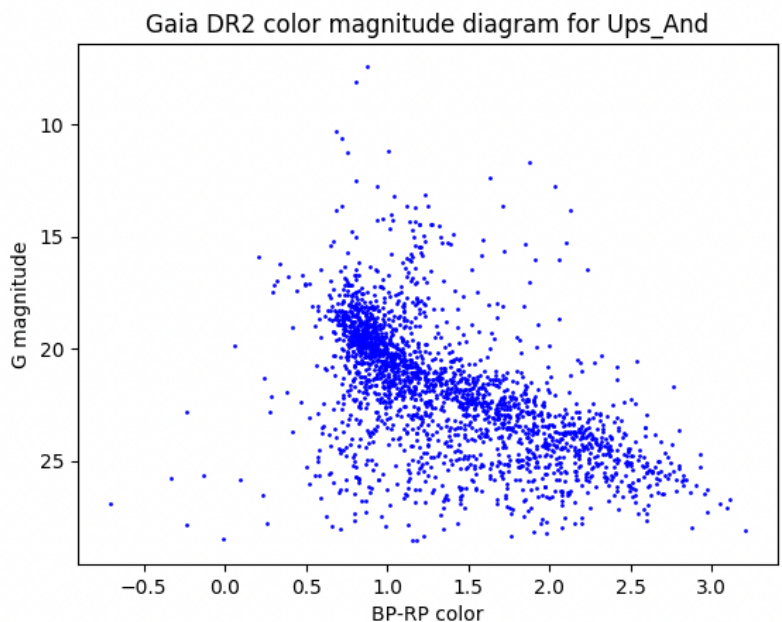

*Figure 4 Color Magnitude Diagram for υ And in Gaia magnitudes*

We also created Besançon simulated catalogs for the target pointings to (a) compare the expected (model) stellar populations to actual data found with Gaia and the Hubble source catalog and (b) extrapolate the derived number density statistics to faint magnitude levels, consistent with anticipated exoplanet brightnesses. The simulations were run with mostly default settings, with the size of the region chosen to ensure ~ 100,000 stars in each simulation and hence negligible Poisson counting uncertainties. (The models for positions near the galactic center therefore covered a smaller area than those with less densely populated areas at higher galactic latitude.) The magnitude and color limits were changed to include stars at all distances and run with a large distance limit of 125kpc to encompass all possible stars. It is also worth noting that the Besancon model also includes a white dwarf and a brown dwarf component.

The number of Gaia sources in each magnitude bin can be plotted and compared to counts expected from the Besançon model to see how closely the models fit the empirical Gaia observations. To convert Besançon Vmag and colors to match the Gaia Gmag values, the following equation, as found in Gaia documentation, was used.

$G = V - 0.0257 - 0.0924 \cdot (V-I) - 0.1623 \cdot (V-I)^2 + 0.0090 \cdot (V-I)^3$

The Besançon model naturally covers a much wider magnitude range than can be observed with Gaia. The Gaia catalog has a magnitude limit of around G mag = 20, while the Besançon model goes to about a magnitude limit of 40, far fainter than required for the present purposes. If the comparison is limited to those regions with overlapping magnitude scales, from approximately 15 to 20, a direct comparison can be shown between the two, as seen in Figure 5.

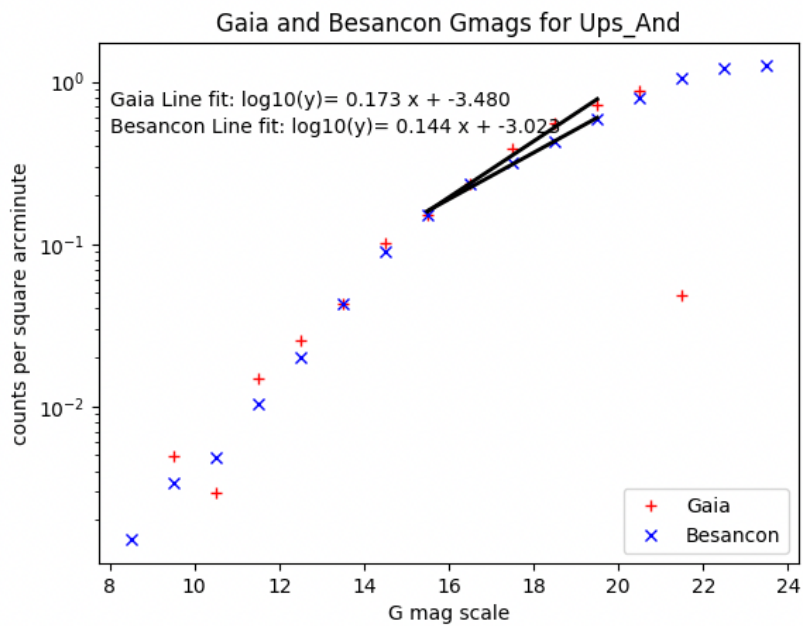

*Figure 5  Gaia counts per magnitude bin per square arcminute compared to Besançon simulations of the same for υ And.*

It is also possible to combine the HSC histogram data of Figure 3 with the Gaia and Besançon catalogs. Since the Hubble data was taken in a redder filter than the Gaia G mag, the Hubble AB magnitudes were plotted against the Gaia magnitudes and an offset was found between them. In order to convert from ACS F775W magnitudes to Gaia G, an offset of 0.15 was added.

Gmag = ACSmag + 0.15

When this is done, it can be seen that there is an excess of sources at the faint end of the scale, around 24 magnitudes and higher. To obtain a grasp on the relative contribution of stars and galaxies (which are not well represented in the Gaia data and are not included in the Besançon model), the concentration index (CI) of the sources listed in the HSC can be used as a filter. This concentration index is defined as the difference in aperture magnitudes using a small and large aperture around the source. The rounder, more point-source-like objects have CI values closer to 1, while more extended sources tend to have higher CI value. Hence, by sorting out the ACS data and plotting the two groups separately, it can be seen that (a) the star like objects are in very good agreement with the Gaia catalog and the Besançon model, and (b) the faint source excess counts seen in the HSC data are very likely due to the emergence of a population of extragalactic sources, recognized by more their more extended CI parameterization, as shown in Figure 6. There were also instances seen in the data where the HSC identified multiple sources on a single target, like a spiral galaxy. Filtering the dataset as described above was an attempt to reject spurious hits, but some may remain in the final catalog, possibly artificially inflating the extended source counts.

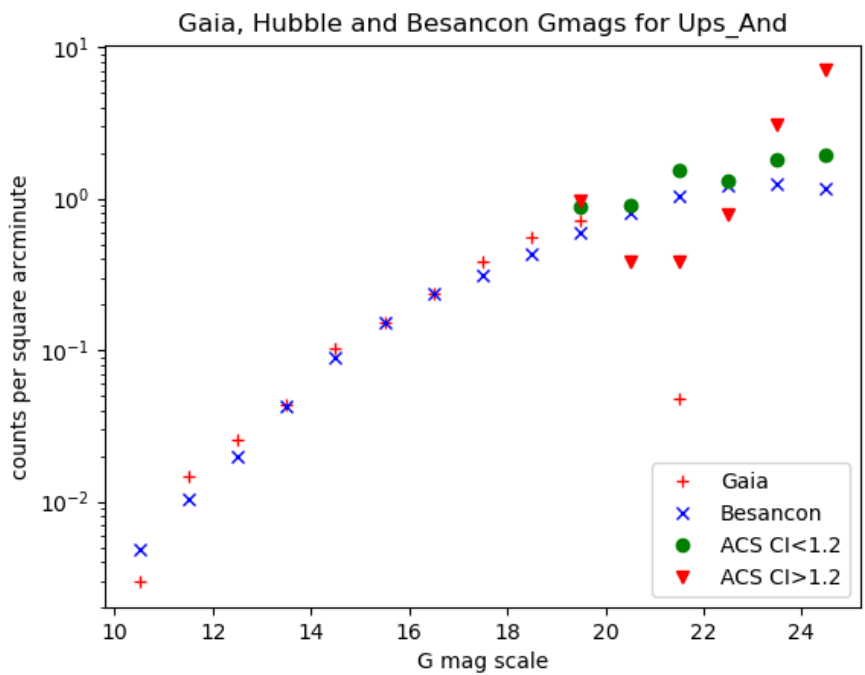

*Figure 6  Comparison of Gaia, ACS and Besançon source counts per magnitudes per square arcminute near Upsilon Andromeda. The ACS data is separated out into point sources (CI < 1.2) and extended sources (CI ≥ 1.2)*

Returning to the specifics of the individual stars near υ And, by using the data in the Gaia catalog (current position and proper motion for all Gaia stars), the 2026 positions of the stars can be examined to determine which stars may be near to the target object in the desired timeframe, as also shown in Figure 1. For this example, the January 1, 2026 position of υ And, as calculated from the SIMBAD J2000 coordinates, is determined to be: 01h 36m 47.4416s +41° 24' 09.7175". The list of stars within one arcminute of that position and their magnitudes as shown in the Gaia DR2 catalog are given in Table 3. The Gaia DR2 observations were in J2015.5 positions, so when the positions were calculated for the year 2026, this put the positions as they would be on July 2,

2026, creating a six-month difference to the SIMBAD position. The star with magnitude 3.89 is υ And with position appropriate to July 2026. This can be seen more clearly in Figure 7. The red plus is the January 1, 2000 position and the red circle is the January 1, 2026 position of υ And. The green × on top of the red circle is from the Gaia catalog calculations of the position, July 2, 2026. The distance calculated in the table below is the distance of the Gaia field star positions from the target star in 2026. (The small difference in calculated positions between Gaia and SIMBAD coordinates, which are about six months apart, result in the distance for the target star not being 0, and could be indicative of the uncertainties in the proper motion calculations as well as the differing dates.)

*Table 3  Positions of stars near Upsilon Andromeda in July 2026.. Position of υ And highlighted in bold italics.. All positions listed are of stars in 2026, based on Gaia positions and proper motions.*

| Source ID | Distance (arcsec) | RA | Dec | G magnitude |
|---|---|---|---|---|
| 348020242217448576 | 55.7656 | 01h36m49.9991s | +41d23m21.9506s | 12.51 |
| *348020448377061376* | *0.2047* | *01h36m47.4364s* | *+41d24m09.5213s* | *3.899* |
| 348020448375872768 | 30.1985 | 01h36m44.9375s | +41d24m20.5898s | 14.891 |
| 348020448375877632 | 36.9352 | 01h36m45.1585s | +41d23m43.1797s | 16.142 |
| 348020448377062528 | 36.7454 | 01h36m45.1633s | +41d23m43.3916s | 15.292 |

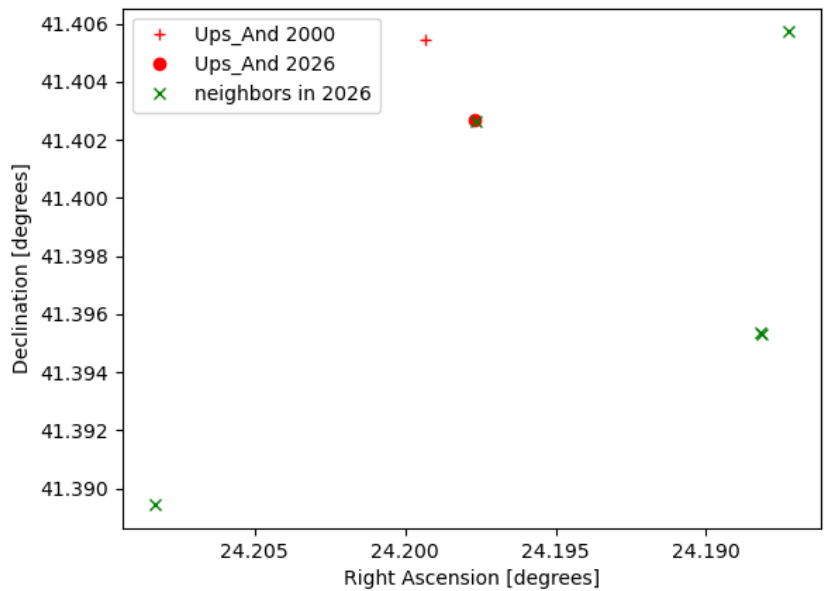

*Figure 7 Position of υ And in 2000 and 2026 with nearby neighbors at 2026 Gaia positions.*

Conclusion for this case study: After examining the area around υ And in Hubble imaging, the Gaia catalog and using the Besançon model, it appears very unlikely that there would be a background object within the CGI apertures that would interfere with any observations. Gaia shows no other cataloged source within the CGI FOV radius (~ 7.2 arcseconds) of the target source.

The widest coronagraphic FOV for Roman CGI is 2.9 arcseconds in diameter, so the two regions we will examine have areas of 6.61 and 40.72 squared arcseconds corresponding to the widest coronagraph FOV and the entire imaged field of view of the CGI.

The equations found by least-square fitting of a straight line to the Gaia data between magnitudes 15 and 20 where the best coverage exists can be used to estimate how many sources might be expected per square arcminute at specific desired magnitudes. (Equations are included within Figure 5 for υ And; similar plots given in the appendices for each target star provide the slopes and intercepts for Table 5.)

Table 5 shows the predicted densities for all targets including υ And in Gaia G mags for magnitude 20 and 24. These magnitudes are approximately 1 and 5 sources per square arcminute, respectively. These values give predicted densities for the apertures ranging from $1.8 \times 10^{-3}$ to $1.1 \times 10^{-2}$ stars in the regions of interest on the low end to $19.2 \times 10^{-3}$ to $5.6 \times 10^{-2}$ stars in the small and large areas of interest on the high end. This shows that there is a very low chance of a source brighter than Gmag of 24 falling by chance within the apertures. The expected probability of 'glints' due to brighter stars outside the field of view is also small, because the number density decreases with increasing star brightness.

4. Statistical Analysis across the target set

By using the methods described for the case study above (υ And) for the sample exploration, similar analysis can be applied to the rest of the dataset. While there is no HSC data for most targets, all targets have data from Gaia and the Besançon model. To compare the Gaia catalog and the Besançon model we obtain a ratio of counts per square arcminute in the overlapping magnitude bins, as shown in Figure 8 for υ And. An average of the ratios between magnitude bins of 15 to 20 can be used as an approximation of how closely the Besançon model matches the observed counts in the Gaia DR2 catalog as shown Table 4 for all targets. The closer this ratio is to 1.0, the closer the Besançon model matches the Gaia catalog. Figure 9 shows this ratio plotted against galactic latitude to demonstrate that Gaia and the Besançon model are in close agreement except at latitudes near zero where small-spatial-scale variability in both extinction and stellar counts may not be well-captured in the Besançon model.

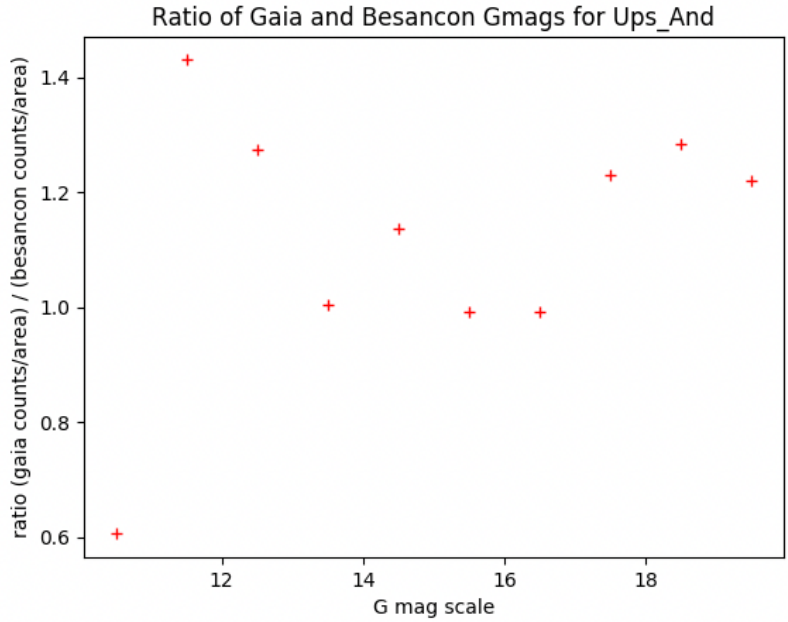

*Figure 8 Ratio of Gaia counts per magnitude bin per square arcminute compared to Besançon model of the same at each magnitude bin for υ And.*

*Table 4 Average ratio of Gaia to Besançon counts per square arcminute for magnitude bins of 15-20.*

| Target | Galactic Latitude (deg) | Ratio of Gaia to Besançon counts |
|---|---|---|
| 14 Her | 46.9449 | 1.018 |
| 47 Uma | 63.3691 | 0.819 |
| 55 Cnc | 37.6989 | 1.148 |
| ε Eri | -48.0513 | 1.052 |
| HD 142 | -66.3866 | 1.022 |
| HD 39091 | -29.7763 | 0.990 |
| HD 114613 | 24.889 | 0.995 |
| HD 134987 | 27.3864 | 0.790 |
| HD 154345 | 37.6635 | 0.979 |
| HD 190360 | -0.6725 | 0.518 |
| HD 192310 | -29.3975 | 0.815 |
| HD 217107 | -53.324 | 1.049 |
| HD 219134 | -3.1986 | 0.787 |
| τ Ceti | -73.4397 | 0.746 |
| υ And | -20.6662 | 1.144 |

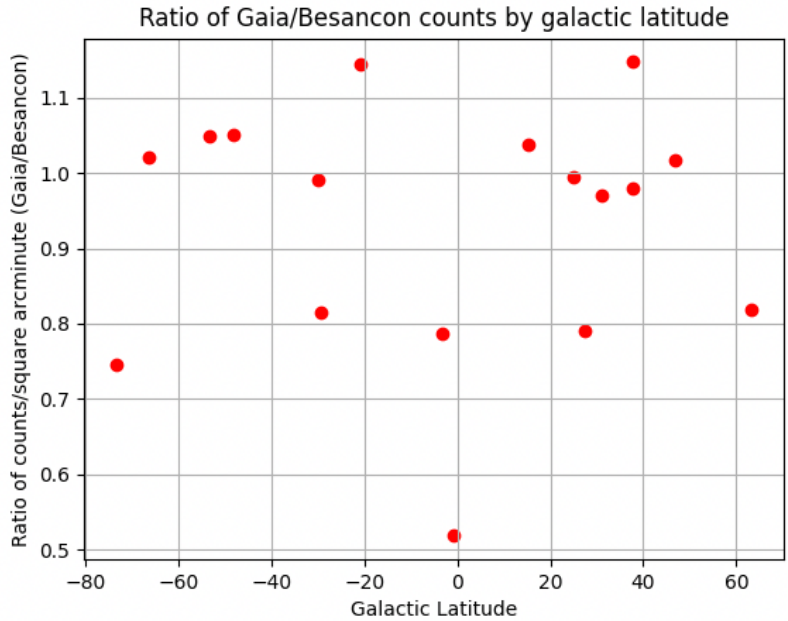

*Figure 9 Ratio of Gaia to Besançon counts per square arcminute (15<G<20) by galactic latitude. The ratio is within 5% of unity for about half of targets, and is no more than 25% off for any target. The exception is HD~190360, which is <1degree from the galactic plane, where small-spatial-scale variability in both extinction and stellar counts may not be well-captured in the Besançon model.*

Table 5 presents for each target, the number of stars found using Gaia in an 18 arcminute radius area around the target stars, the slope and intercept of the fitted lines as shown in Figure 5 and other figures for each target, and the expected number of objects at magnitudes of 20 and 24 per square arcminute. The magnitude bins cover integer ranges of magnitudes, so the magnitude 20 bin covers all magnitudes between 20 and 21, and the magnitude bin of 24 covers all magnitudes between 24 and 25. To get the expected number of sources within those bins, the following equation applies:

log10(number of sources per square arcminute) = Slope * magnitude + Intercept, keeping in mind this is a log relationship.

Table 5 *Gaia statistics for target stars. The slope and intercept values are those shown on the Gaia/Besançon plots with magnitudes vs. star counts per square arcminute. The last column is from the Besançon model, using the Besançon slope and intercept from the individual plots, just as a comparison to the predicted Gaia numbers at mag 24.*

| Target | Gaia sources | Gaia Slope | Gaia Intercept | Gaia Source density at mag 20 | Gaia Source density at mag 24 | Besançon source density at mag 24 |
|---|---|---|---|---|---|---|
| 14 Her | 1460 | 0.180 | -4.012 | 0.389 | 2.046 | 1.663 |
| 47 Uma | 657 | 0.172 | -4.202 | 0.175 | 0.855 | 0.796 |
| 55 Cnc | 1242 | 0.145 | -3.455 | 0.284 | 1.085 | 0.914 |
| ε Eri | 926 | 0.1797 | -4.196 | 0.251 | 1.312 | 0.764 |
| HD 142 | 1069 | 0.162 | -3.848 | 0.250 | 1.117 | 1.270 |
| HD 39091 | 3178 | 0.183 | -3.701 | 0.919 | 4.972 | 4.256 |
| HD 114613 | 5271 | 0.202 | -3.817 | 1.695 | 10.925 | 9.795 |
| HD 134987 | 5553 | 0.207 | -3.871 | 1.811 | 12.129 | 21.878 |
| HD 154345 | 1804 | 0.168 | -3.667 | 0.486 | 2.277 | 2.366 |
| HD 190360 | 49024 | 0.273 | -4.272 | 15.334 | 189.304 | 277.971 |
| HD 192310 | 5588 | 0.208 | -3.911 | 1.738 | 11.754 | 19.815 |
| HD 217107 | 1072 | 0.148 | -3.506 | 0.282 | 1.101 | 1.422 |
| HD 219134 | 20601 | 0.236 | -3.871 | 6.964 | 61.045 | 88.920 |
| τ Ceti | 685 | 0.1996 | -4.732 | 0.182 | 1.1419 | 0.820 |
| υ And | 3225 | 0.173 | -3.480 | 0.955 | 4.696 | 2.710 |

Table 6 *Using the values of stellar density per arcminute squared in the previous table, the number of sources which would be expected to fall in the Roman CGI full field of view are given for two different magnitudes.*

| Target | Predicted Gaia mag 20 stars in CGI FOV | Predicted Gaia mag 24 stars in CGI FOV | Predicted Besançon mag 20 stars in CGI FOV | Predicted Besançon mag 24 stars in CGI FOV |
|---|---|---|---|---|
| 14 Her | 4.40E-03 | 2.31E-02 | 4.00E-03 | 1.88E-02 |
| 47 Uma | 1.98E-03 | 9.69E-03 | 2.18E-03 | 9.00E-03 |
| 55 Cnc | 3.21E-03 | 1.22E-02 | 2.77E-03 | 1.03E-02 |
| ε Eri | 2.83E-03 | 1.48E-02 | 2.25E-03 | 8.64E-03 |
| HD 142 | 2.84E-03 | 1.27E-02 | 2.94E-03 | 1.43E-02 |
| HD 39091 | 1.04E-02 | 5.61E-02 | 9.96E-03 | 4.81E-02 |
| HD 114613 | 1.91E-02 | 1.23E-01 | 1.84E-02 | 1.11E-01 |
| HD 134987 | 2.05E-02 | 1.37E-01 | 2.97E-02 | 2.47E-01 |
| HD 154345 | 5.50E-03 | 2.58E-02 | 5.59E-03 | 2.68E-02 |
| HD 190360 | 1.74E-01 | 2.14E+00 | 3.03E-01 | 3.14E+00 |
| HD 192310 | 1.96E-02 | 1.33E-01 | 2.74E-02 | 2.24E-01 |
| HD 217107 | 3.19E-03 | 1.24E-02 | 3.46E-03 | 1.61E-02 |
| HD 219134 | 7.88E-02 | 6.90E-01 | 1.04E-01 | 1.01E+00 |
| τ Ceti | 2.06E-03 | 1.29E-02 | 2.23E-03 | 9.28E-03 |
| υ And | 1.08E-02 | 5.31E-02 | 8.14E-03 | 3.07E-02 |

5. Conclusion

As will be seen in the analysis of the individual targets in Appendix A, Gaia shows few stars within the radius of the full CGI FOV (~ 7.2 arcseconds) of the target stars. In Table 6, the stellar densities in the full CGI field of view for magnitude 20 and 24 objects as expected from both Gaia and the Besançon models are shown. The Besançon value is a projection from the overlap region following the Gaia procedure, for comparison – since these projections invariably reside above the actual model at fainter magnitudes, the estimated density and hence contamination probability is conservative in overestimating the stellar density. These densities show that for most fields it is very unlikely that an object would randomly fall within the Roman CGI field of view.

The low Galactic latitude star HD 190360 shows that there are predicted to be around 2-3 stars of magnitude 24 in the full field of view, but if we consider only the smaller high contrast region, (radius 2.9 arcseconds) Gaia shows 0.348 stars of mag 24 and 0.028 stars of mag 20 in this region. This shows that there are not likely to be contaminating sources nearby enough to be a problem for the great majority of the expected observations, though at the lowest Galactic latitude, faint stellar background sources can be a concern.

The contamination analysis is based primarily on stellar statistics; however as seen in Figure 6, Figure 16, Figure 20, and Figure 43, (the stars with HSC data), a population of extragalactic sources is evident at the faintest relevant magnitudes. An effort was made to remove any spurious hits from the HSC catalogs, but there could still be a few such cases in the catalogs. Such spurious hits would contribute to, but not likely dominate the rise seen in the extragalactic sources. Modulo cosmic variance, the number of such sources is likely to be independent of galactic latitude and comparable to order of magnitude to the density of sources near these four stars. This density corresponds to an average of approximately 8 counts per square arcminute, which gives approximately one percent chance of finding an extragalactic background source within the Roman CGI high contrast region. These counts are loosely consistent with the galaxy counts found in previous work, e.g. Eliche-Moral 2006. This is on the verge of becoming significant, but is unlikely to be a major problem even at these bright magnitudes.

Acknowledgements

Hubble Source Catalog: Based on observations made with the NASA/ESA Hubble Space Telescope, and obtained from the Hubble Legacy Archive, which is a collaboration between the Space Telescope Science Institute (STScI/NASA), the Space Telescope European Coordinating Facility (ST-ECF/ESAC/ESA) and the Canadian Astronomy Data Centre (CADC/NRC/CSA).

This work has made use of data from the European Space Agency (ESA) mission Gaia (https://www. cosmos.esa.int/gaia), processed by the Gaia Data Processing and Analysis Consortium (DPAC, https://www.cosmos.esa.int/web/gaia/dpac/consortium). Funding for the DPAC has been provided by national institutions, in particular the institutions participating in the Gaia Multilateral Agreement.

*This research has made use of the Besançon Galaxy Model,*
*found here: https://model.obs-besancon.fr/modele_home.php*

*This research has made use of the SIMBAD database,*
*operated at CDS, Strasbourg, France.*
2000,A&AS,143,9 , "The SIMBAD astronomical database", Wenger et al.

Kasdin, N. et al. 2020, 'The Nancy Grace Roman Space Telescope Coronagraph Instrument (CGI) technology demonstration', Kasdin N. et al. *SPIE 11443,* SPIE Astronomical Telescopes and Instrumentation, 2020, Online Only, 15 December 2020.

Eliche-Moral, M. C., Balchells, M., Prieto, M., Garcia-Dabo, C. E., Erwin, P., Cristobal-Hornillos, D., 2006, 'Goya Survey: U and B Number counts in the Groth Westphal Strip'. APJ 639:644-671

APPENDIX

A. Results for full target list
The appendix here shows results, plots and tables from an analysis of Gaia data and the Besançon model, and Hubble data where it is available, for the targets other than Upsilon Andromeda.

A.1. 14 Hercules

There is no Hubble data for 14 Hercules, so the analysis only includes the Gaia catalog and Besançon models. The first plot shows the comparison of the Gaia catalog to the Besançon models, showing source counts per square arcminute broken out by Gaia G magnitudes. The second figure and the table show the star positions from Gaia within one arcminute of the target star in 2026.

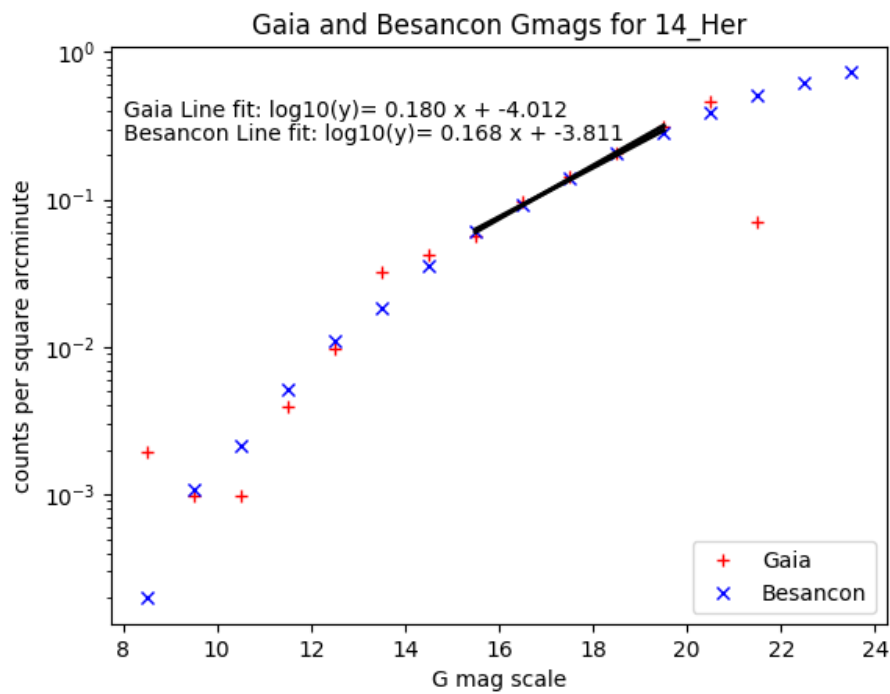

*Figure 10 Gaia counts per magnitude bin per square arcminute compared to Besançon simulations of the same for 14 Her.*

14 Her in 2026 is at RA, Dec of 16h10m24.631s +43d48m55.8183s as calculated from coordinates and proper motions found in SIMBAD in J2000 coordinates.

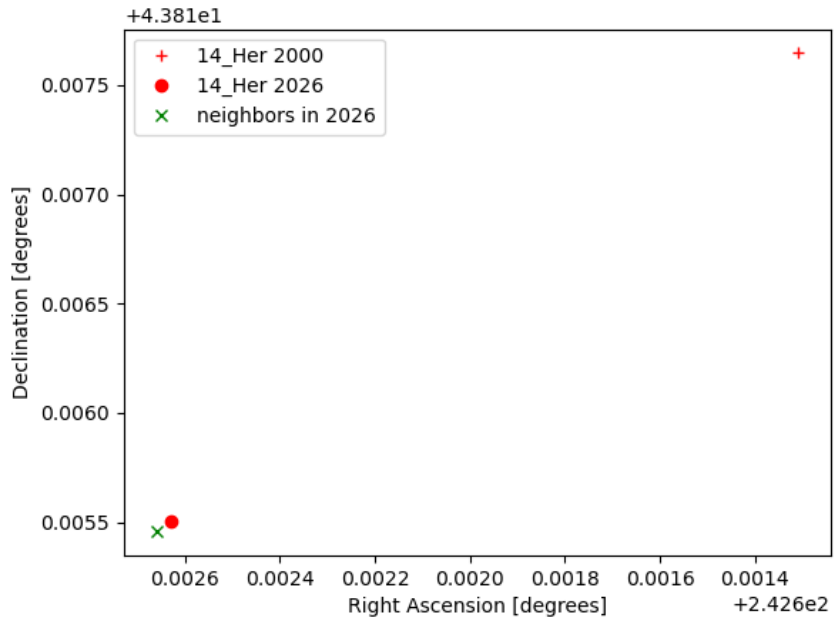

*Figure 11 Positions of 14 Her in 2000 and 2026 compared with positions of other Gaia sources within one arcminute.*

*Table 7 Positions of stars within one arcminute of 14 Her in 2026. Only the star itself is within 1 arcmin.*

| Source ID | Distance (arcsec) | RA | Dec | G mag |
|---|---|---|---|---|
| *1385293808145621504* | *0.19393* | *16h10m24.6385s* | *+43d48m55.6424s* | *6.3793* |

A.2 47 Ursa Majoris

There is no Hubble data for 47 Ursa Majoris, so the analysis only includes the Gaia catalog and Besançon models. The first plot shows the comparison of the Gaia catalog to the Besançon models, showing source counts per square arcminute broken out by Gaia G magnitudes. The second figure and the table show the star positions from Gaia within one arcminute of the target star in 2026.

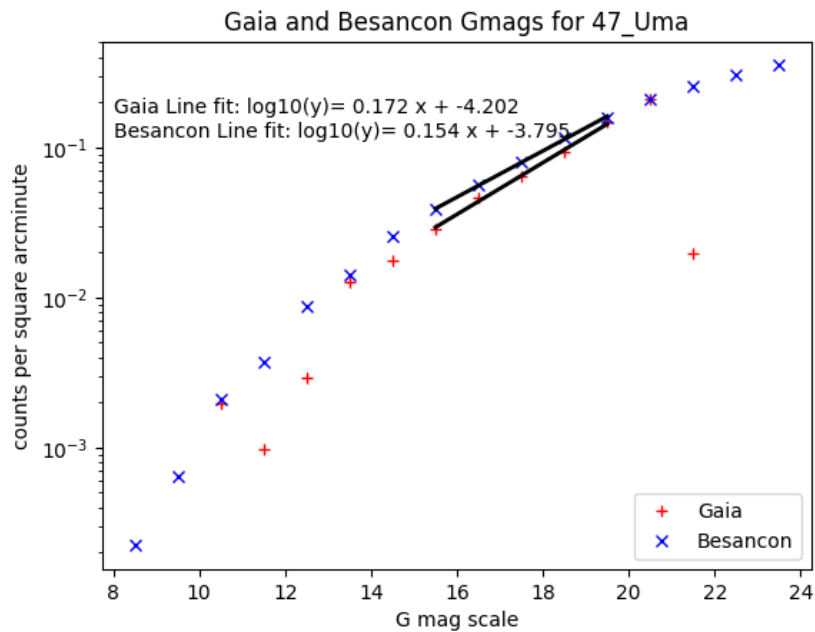

Figure 12 *Gaia counts per magnitude bin per square arcminute compared to Besançon simulations of the same for 47 UMa.*

The position of 47 UMa in 2026 in RA, Dec is 10h59m27.2495s +40d25m50.3504s as calculated from coordinates and proper motions found in SIMBAD in J2000 coordinates.

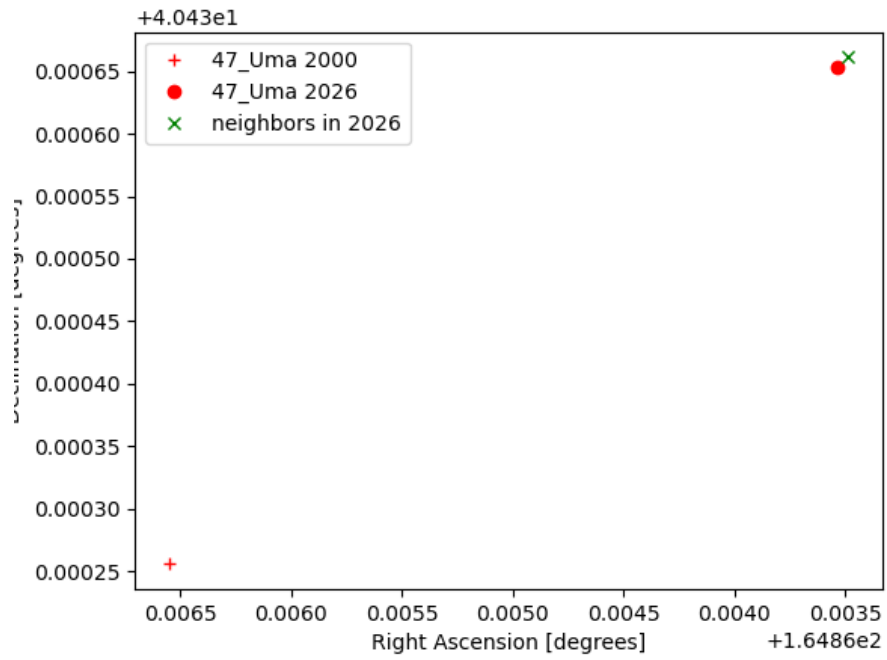

*Figure 13 Positions of 47 UMa in 2000 and 2026 compared with positions of other Gaia sources within one arcminute.*

*Table 8 Positions of stars within one arcminute of 47 UMa in 2026. Only the star itself is within 1 arcmin.*

| Source ID | Distance (arcseconds) | RA | Dec | G mag |
|---|---|---|---|---|
| *777254360337133312* | *0.14988* | *10h59m27.2366s* | *+40d25m50.3803s* | *4.8306* |

A.3 55 Cancri

There is Hubble data for 55 Cancri, images both on the star and in parallel fields. The proper motion of the star can be shown in a single WFC3 IR image centered on the star. This image shown below on the left is from 2015, ico517siq_drz.fits. The green circle shows the position of the star in 2000, while the red circle shows where the star will be in 2026.

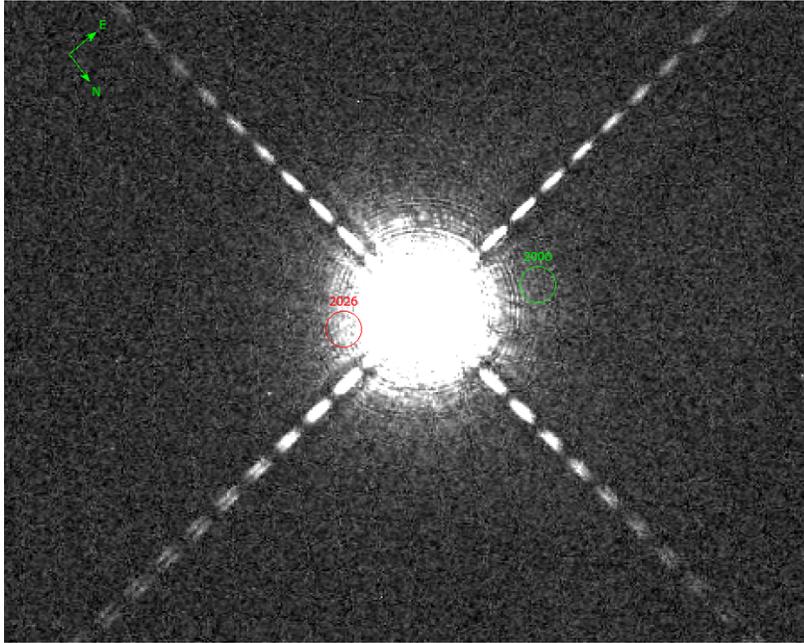

*Figure 14 Hubble WFC3 IR image of 55 Cancri taken in 2015. The circle in green shows the 2000 position, while the circle in red shows where the star will be in 2026.*

There are several parallel fields taken within 18 arcminutes of the target star. There are three WFPC2 F606W fields, and one ACS F775W image. Two of those WFPC2 images and the ACS fields combine only two images to have exposure times of 2000 seconds or less, while one WFPC2 image consists of 30 individual combined images with a total exposure time of 31553 seconds. Looking at this combined image in the Hubble Source catalog shows 128 sources total, which gives a total of approximately 22 sources per square arcminute in the combined image, since a WFPC2 image has a 5.692 square arcminute field of view. If we divide this out using the concentration index (CI) value to separate possible stars from probable galaxies with CI values < 1.2 being more likely to be stars and CI values > 1.2 more likely to be galaxies (as was done in the analysis above, the breakdown is 9 sources with a CI less than 1.2 and 119 with a CI greater than 19.

Table 9 Comparison of total sources in WFPC2 combined image (total and broken out by CI value) compared to the Gaia catalog for the area around 55 Cancri.

| Catalog being used | Area of coverage (square arcsec) | Total number of sources | Source density per square arcminute |
|---|---|---|---|
| Total of WFPC2 | 5.7 | 128 | 22.5 |
| Sources with CI < 1.2 | 5.7 | 9 | 1.6 |
| Sources with CI > 1.2 | 5.7 | 119 | 20.9 |
| Gaia (within r=18 arcmin) | 1017.9 | 1242 | 1.2 |
| Besançon | 25920.0 | 124109 | 4.8 |

This table shows that the Hubble catalog contains many more sources than the Gaia catalog, but also that most of the sources found in this particular field of view are more likely to be galaxies than stars. If only stars are compared, then Gaia and the Hubble Source catalog are in closer agreement.

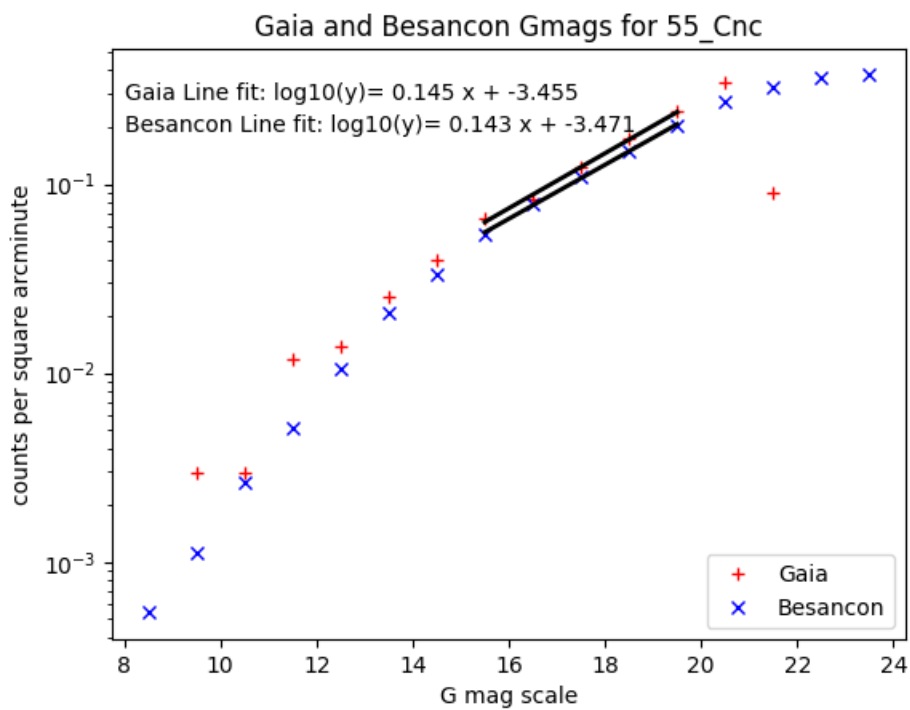

Figure 15 Gaia counts per magnitude bin per square arcminute compared to Besançon simulations of the same for 55 Cnc.

Adding in the WFPC2 data described above to Figure 14 gives us Figure 15, which shows that the Hubble WFPC2 magnitudes for the stars (CI< 1.2) are well aligned with Gaia and the Besançon catalogs, while the galaxies dominate at the faintest magnitudes. To convert from WFPC2 F606W magnitudes to Gaia G magnitudes, an offset of 0.8 was subtracted.

Gmag = WFPC2mag – 0.8

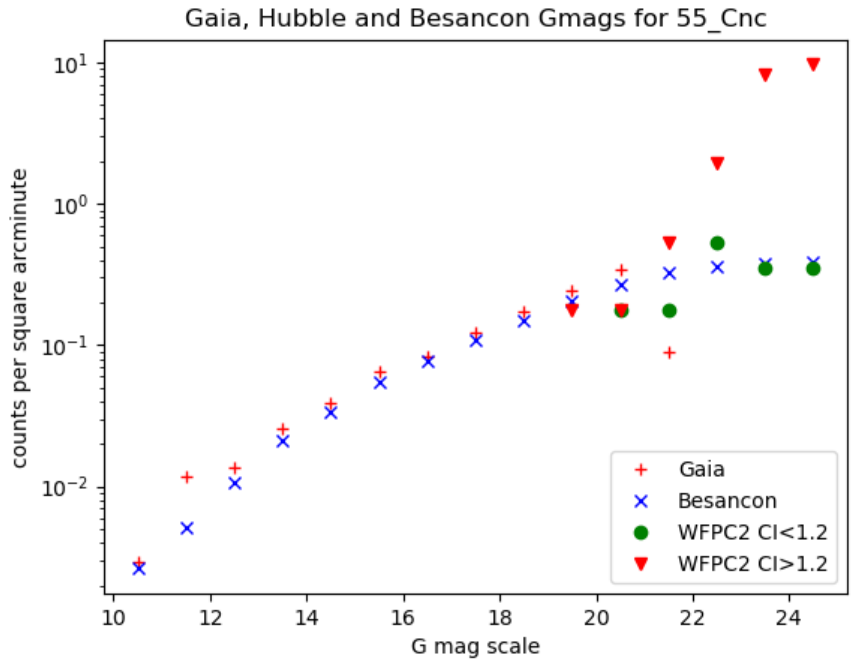

*Figure 16 Comparison of Gaia, WFPC2 and Besançon source counts per magnitudes per square arcminute near 55 Cancri. The WFPC2 data is separated out into point sources (CI < 1.2) and extended sources (CI ≥ 1.2)*

The position of 55 Cnc in 2026 in RA, Dec is 08h52m34.8542s +28d19m44.876s as calculated from coordinates and proper motions found in SIMBAD in J2000 coordinates.

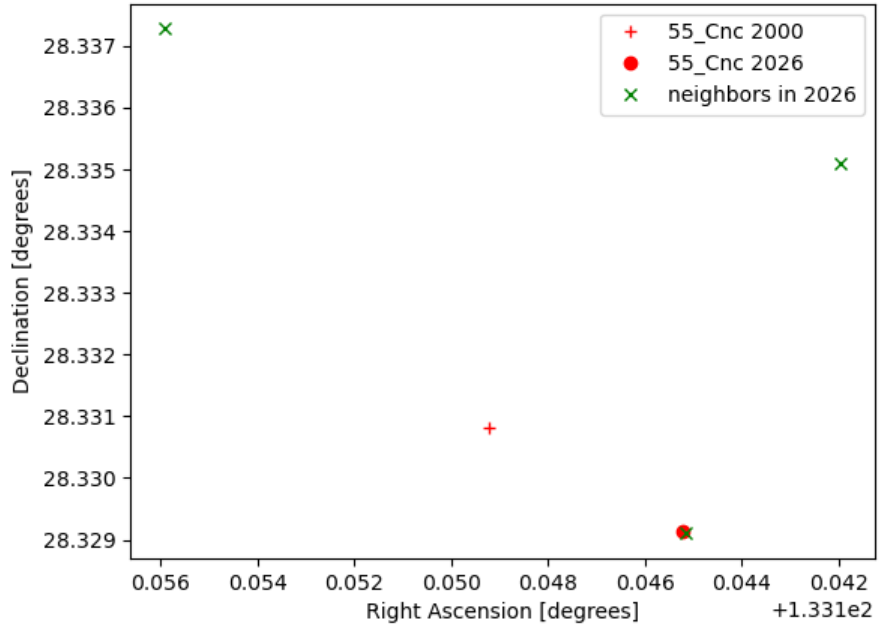

*Figure 17 Positions of 55 Cnc in 2000 and 2026 compared with positions of other Gaia sources within one arcminute.*

*Table 10 Positions of stars within one arcminute of 55 Cnc in 2026. Target star in bold italics.*

| Source ID | Distance | RA | Dec | G mag |
|---|---|---|---|---|
| ***704967037090946688*** | ***0.26233*** | ***08h52m34.8362s*** | ***+28d19m44.7652s*** | ***5.7144*** |
| 704967037089999232 | 23.8737 | 08h52m34.0675s | +28d20m06.3719s | 20.6613 |
| 704967140168773888 | 44.8516 | 08h52m37.4243s | +28d20m14.204s | 19.2226 |

A.4 Epsilon Eridani

Epsilon Eridani also has data taken with Hubble, which allows us to use both the Hubble Source Catalog and the Hubble Legacy archive to examine images of the source itself and the parallel fields around it to see how well the statistics from Hubble match those of Gaia and the Besançon simulated catalog.

The image below is a 1996 WFPC2/PC combined image from the HLA, hst_0683_02_wfpc2_total_wf_sci.fits, centered on the target. The green circle indicates the positions of the star in 2000 and the red circle indicates the position in 2026.

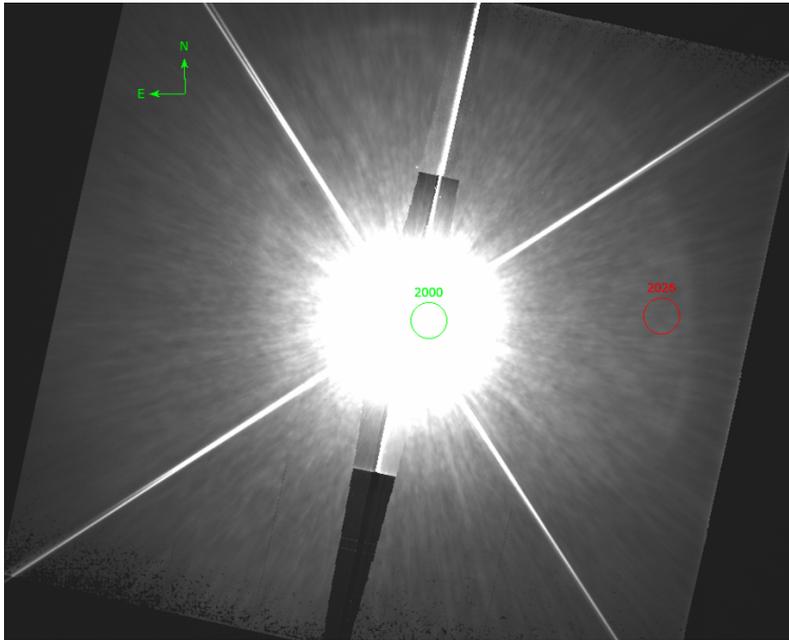

*Figure 18 Hubble WFPC2/PC HLA image of ε Eri taken in 1996. The circle in green shows the 2000 position, while the circle in red shows where the star will be in 2026.*

A wider shot with the central region washed out by the bright star shows the surrounding area with one or two fainter stars and several background galaxies.

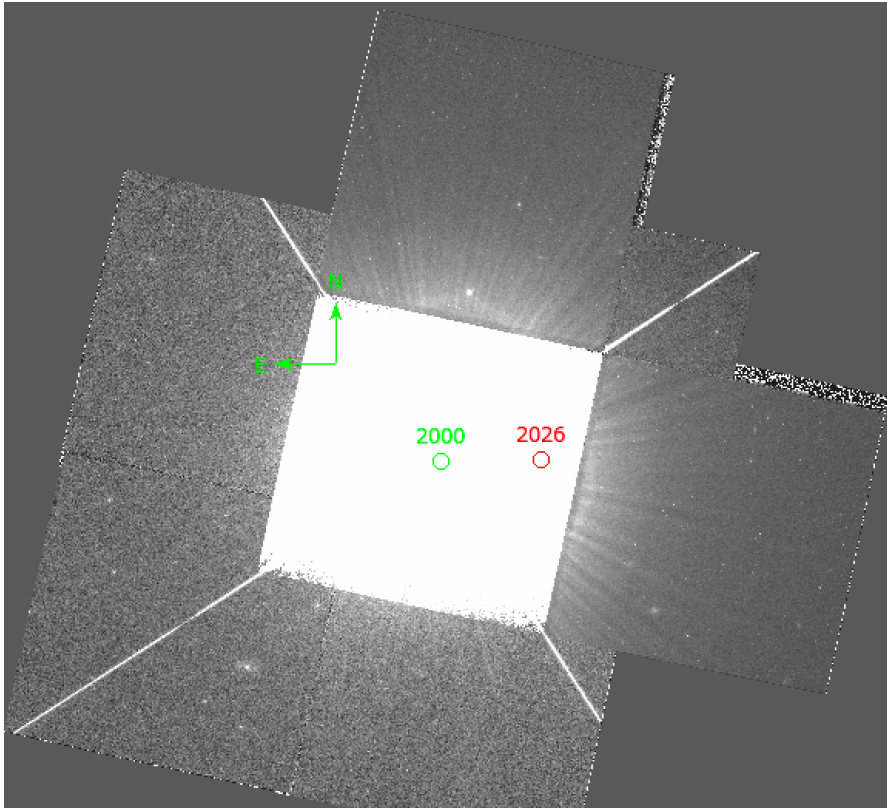

*Figure 19 Hubble WFPC2/PC HLA image of ε Eri taken in 1996. The circle in green shows the 2000 position, while the circle in red shows where the star will be in 2026. The wider field of view shows the background galaxies and one or two fainter stars.*

In the Hubble Legacy archive, there are two ACS F775W images within the 18 arcminute radius. Using the Hubble source catalog, we can obtain a catalog of sources.

By limiting the sources to only those found in more than one ACS image, there were 905 stars found in the two ACS images. While there are some known spurious sources (multiple sources counted in the arms of a large spiral galaxy for instance), this catalog can be compared to what was found with Gaia and predicted with the Besançon simulation. There are a large number of galaxies in this image, which is shown when the sources are broken out by concentration index. CI values less than 1.2 are assumed to be point sources, while CI values greater than 1.2 are assumed to be more elongated and likely galaxies.

Table 11 *Comparison of total sources in two fields of ACS combined images (total and broken out by CI value) compared to the Gaia catalog for the area around ε Eri.*

| Catalog being used | Area of coverage (square arcsec) | Total number of sources | Source density per square arcminute |
|---|---|---|---|
| Total of ACS | 22.7 | 905 | 40 |
| Sources with CI < 1.2 | 22.7 | 92 | 4.1 |
| Sources with CI > 1.2 | 22.7 | 813 | 35.9 |
| Gaia | 1017.9 | 926 | 0.9 |
| Besançon | 33120 | 137986 | 4.2 |

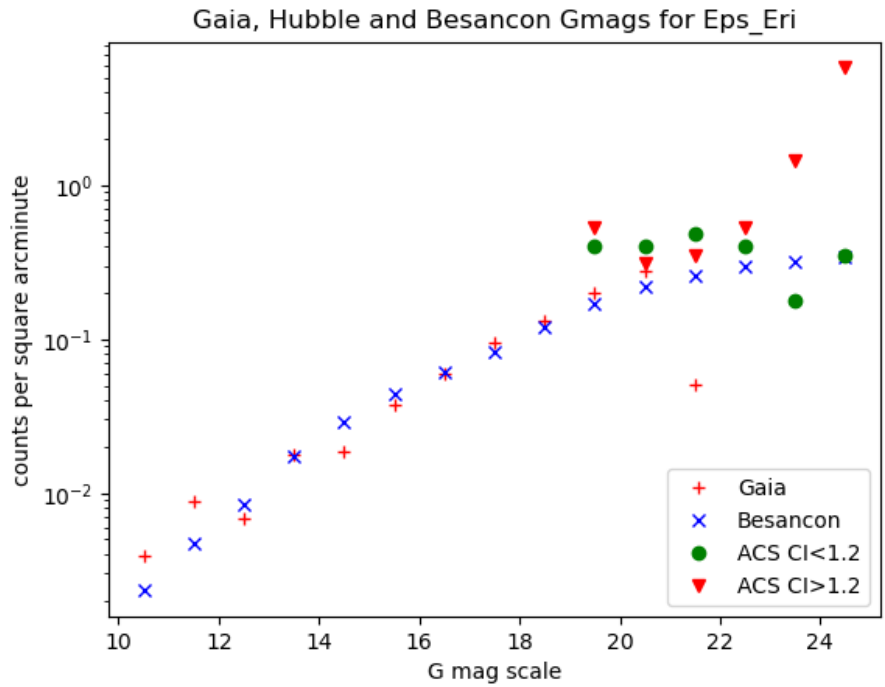

Figure 20 *Comparison of Gaia, ACS and Besançon source counts per magnitudes per square arcminute near ε Eri. The ACS data is separated out into point sources (CI < 1.2) and extended sources (CI ≥ 1.2).*

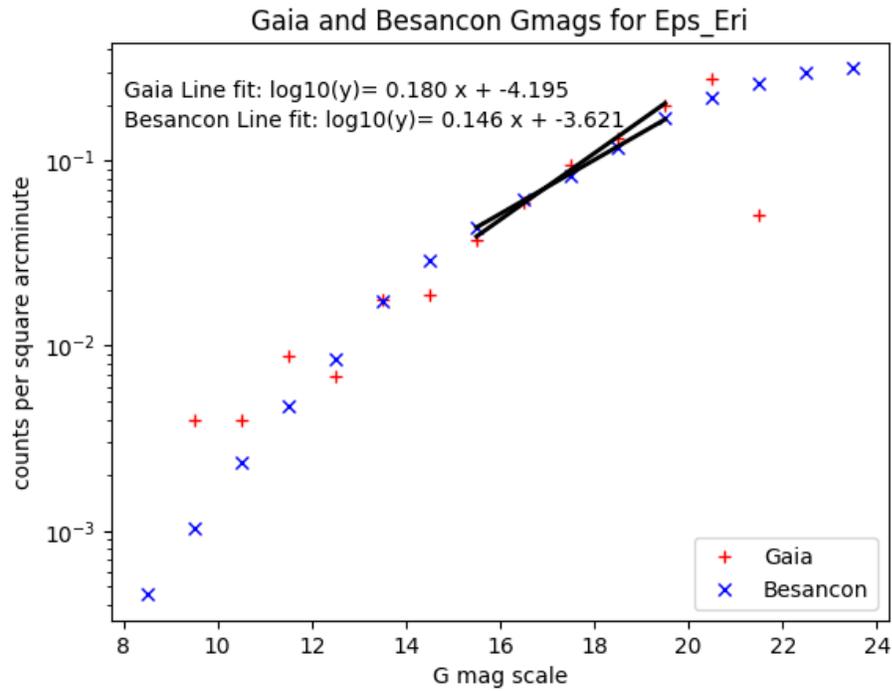

*Figure 21 Gaia counts per magnitude bin per square arcminute compared to Besançon simulations of the same for ε Eri.*

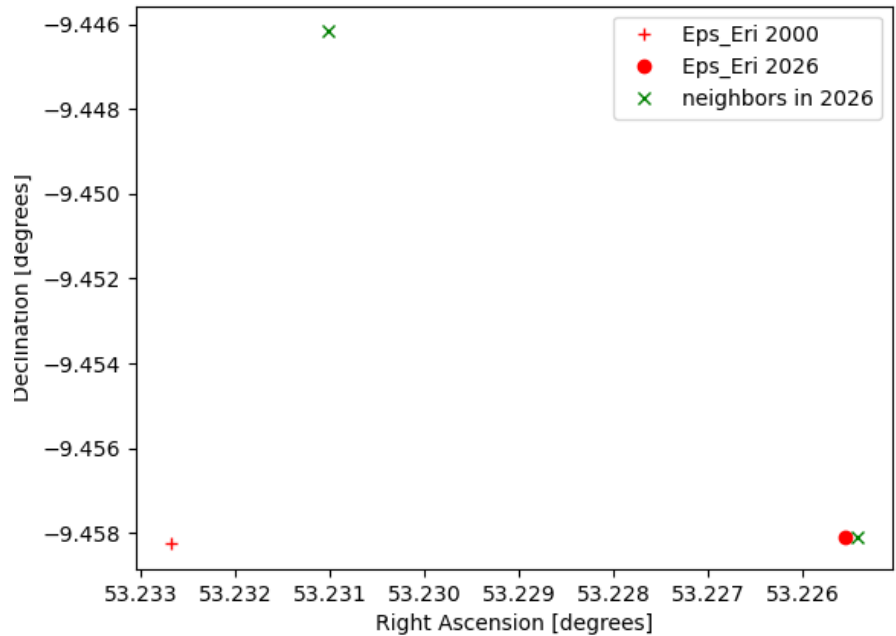

*Figure 22 Positions of ε Eri in 2000 and 2026 compared with positions of other Gaia sources within one arcminute.*

*Table 12 Positions of stars within one arcminute of ε Eri in 2026. Target star in bold italics.*

| Source ID | Dist | RA | Dec | G mag |
|---|---|---|---|---|
| ***5164707970261630080*** | ***0.49153*** | ***03h32m54.0982s*** | ***-09d27m29.1945s*** | ***3.3691*** |
| 5164708142060320768 | 47.18 | 03h32m55.4431s | -09d26m46.2211s | 17.628 |

A.5 HD 142

There is no Hubble data for HD 142, so the analysis only includes the Gaia catalog and Besançon models. The first plot shows the comparison of the Gaia catalog to the Besançon models, showing source counts per square arcminute broken out by Gaia G magnitudes. The second figure and the table show the star positions from Gaia within one arcminute of the target star in 2026.

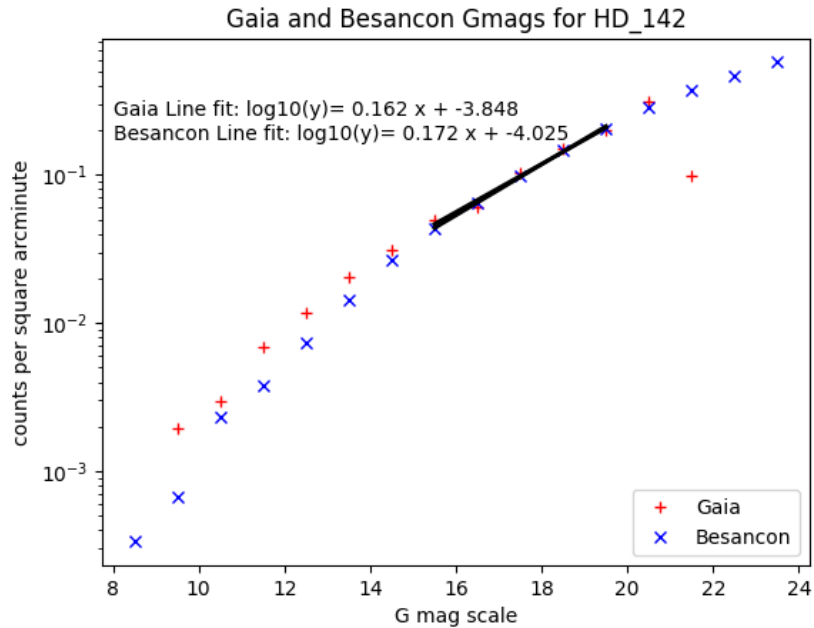

*Figure 23 Gaia counts per magnitude bin per square arcminute compared to Besançon simulations of the same for HD 142.*

The position of HD 142 in 2026 in RA, Dec is 00h06m20.6977s -49d04m31.7s as calculated from coordinates and proper motions found in SIMBAD in J2000 coordinates.

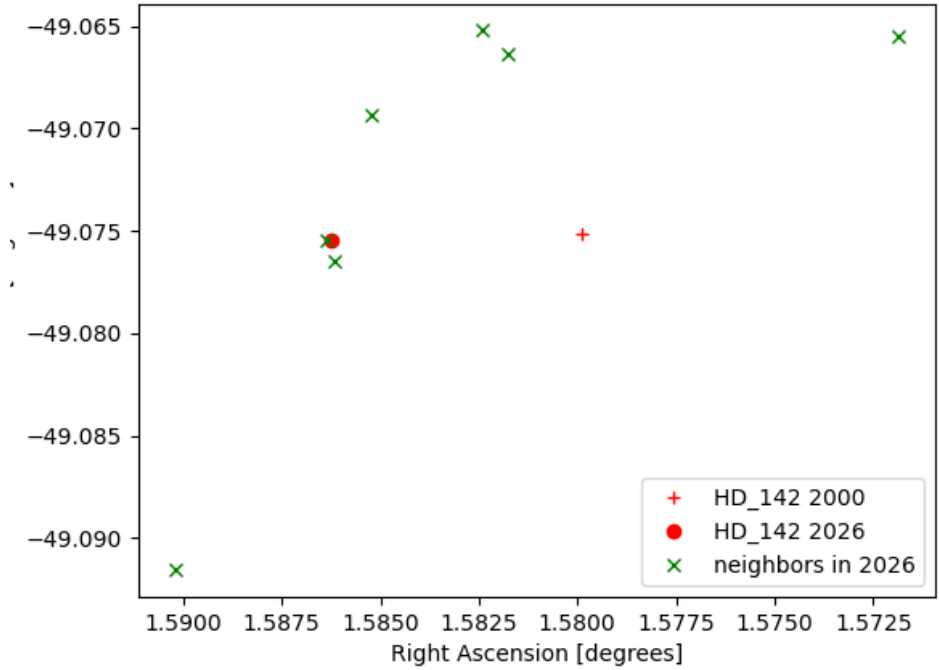

*Figure 24 Positions of HD 142 in 2000 and 2026 compared with positions of other Gaia sources within one arcminute.*

*Table 13 Positions of stars within one arcminute of HD 142 in 2026. Target star in bold italics.*

| Source ID | Distance (arcseconds) | RA | Dec | G mag |
|---|---|---|---|---|
| 4976894891564743168 | 58.744 | 00h06m21.6513s | -49d05m29.6921s | 13.4579 |
| *4976894960284258048* | *0.2854* | *00h06m20.7262s* | *-49d04m31.7525s* | *5.5647* |
| 4976894960284258176 | 3.67068 | 00h06m20.677s | -49d04m35.365s | 10.5334 |
| 4976895166441959808 | 34.4445 | 00h06m19.6275s | -49d03m58.9003s | 19.8742 |
| 4976895166442687744 | 22.0543 | 00h06m20.4566s | -49d04m09.7733s | 18.9162 |
| 4976895162145914368 | 37.9892 | 00h06m19.7818s | -49d03m54.7925s | 12.304 |
| 4976895235160081792 | 49.4881 | 00h06m17.2428s | -49d03m55.6944s | 20.5782 |

A.6 HD 39091

There is no Hubble data for HD 39091, so the analysis only includes the Gaia catalog and Besançon models. The first plot shows the comparison of the Gaia catalog to the Besançon models, showing source counts per square arcminute broken out by Gaia G magnitudes. The second figure and the table show the star positions from Gaia within one arcminute of the target star in 2026.

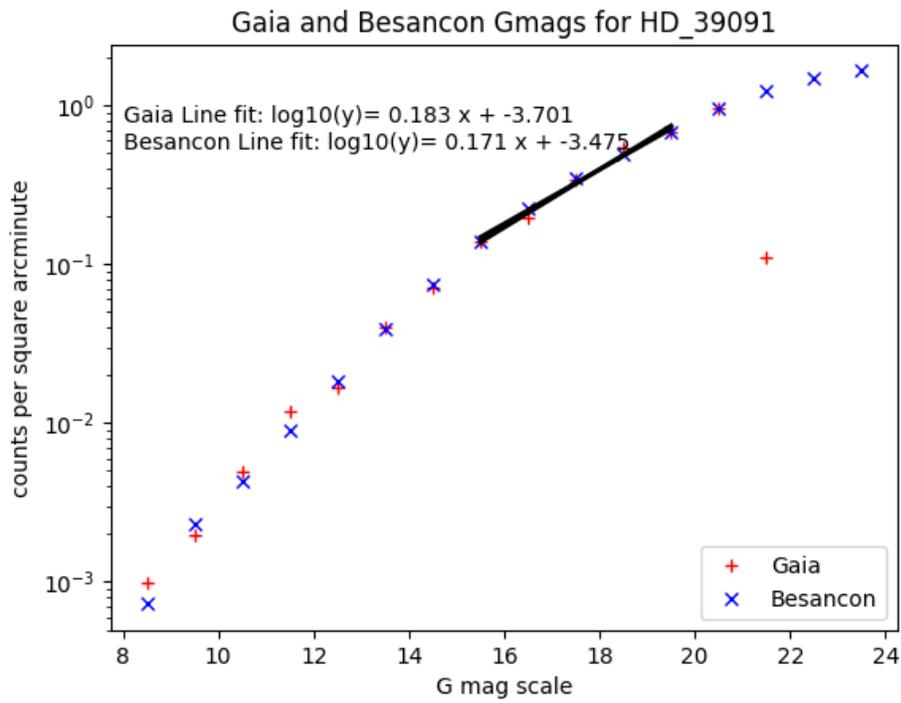

*Figure 25 Gaia counts per magnitude bin per square arcminute compared to Besançon simulations of the same for HD 39091.*

The position of HD 39091 in 2026 in RA, Dec is 05h37m13.1469s -80d27m41.566s as calculated from coordinates and proper motions found in SIMBAD in J2000 coordinates.

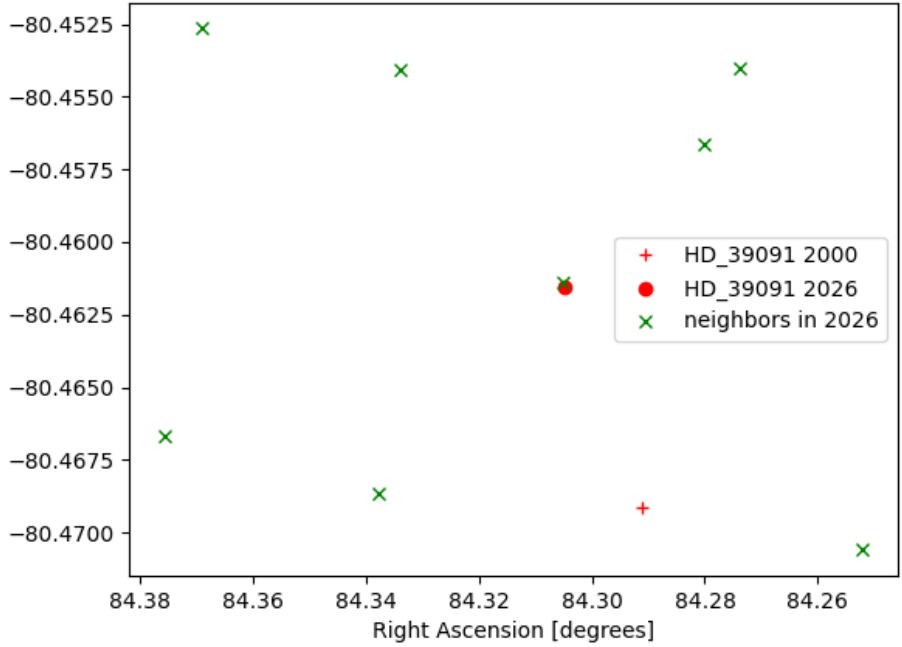

*Figure 26 Positions of HD 39091 in 2000 and 2026 compared with positions of other Gaia sources within one arcminute.*

*Table 14 Positions of stars within one arcminute of HD 39091 in 2026. Target star in bold italics.*

| Source ID | Distance (arcseconds) | RA | Dec | G mag |
|---|---|---|---|---|
| 4623036934091503232 | 22.9469 | 05h37m07.2486s | -80d27m23.9164s | 18.3499 |
| 4623036899732204032 | 45.3085 | 05h37m00.4829s | -80d28m14.1685s | 20.6984 |
| 4623036865373916160 | 32.2556 | 05h37m21.0681s | -80d28m07.1206s | 19.6821 |
| 4623036934091936640 | 32.7732 | 05h37m05.6841s | -80d27m14.5526s | 20.7209 |
| *4623036865373793408* | *0.54777* | *05h37m13.2043s* | *-80d27m41.0371s* | *5.4907* |
| 4623036865372029312 | 46.1935 | 05h37m30.1729s | -80d28m00.1066s | 20.049 |
| 4623036968453008640 | 31.949 | 05h37m20.1274s | -80d27m14.7436s | 17.6043 |
| 4623036964157027072 | 49.9753 | 05h37m28.5763s | -80d27m09.5457s | 18.3 |

A.7 HD 114613

There is no Hubble data for HD 114613, so the analysis only includes the Gaia catalog and Besançon models. The first plot shows the comparison of the Gaia catalog to the Besançon models, showing source counts per square arcminute broken out by Gaia G magnitudes. The second figure and the table show the star positions from Gaia within one arcminute of the target star in 2026.

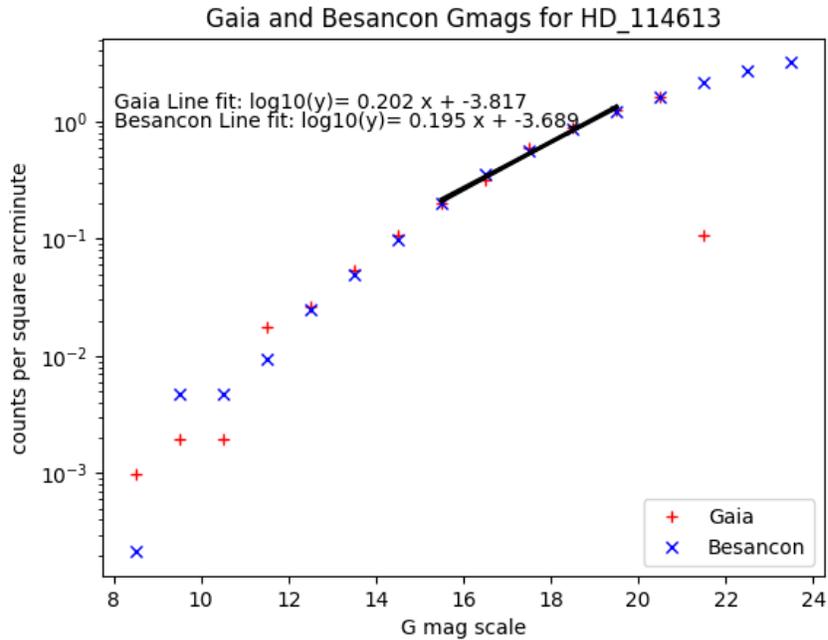

*Figure 27 Gaia counts per magnitude bin per square arcminute compared to Besançon simulations of the same for HD 114613.*

The position of HD 114613 in 2026 in RA, Dec is 13h12m02.3478s -37d48m09.6915s as calculated from coordinates and proper motions found in SIMBAD in J2000 coordinates.

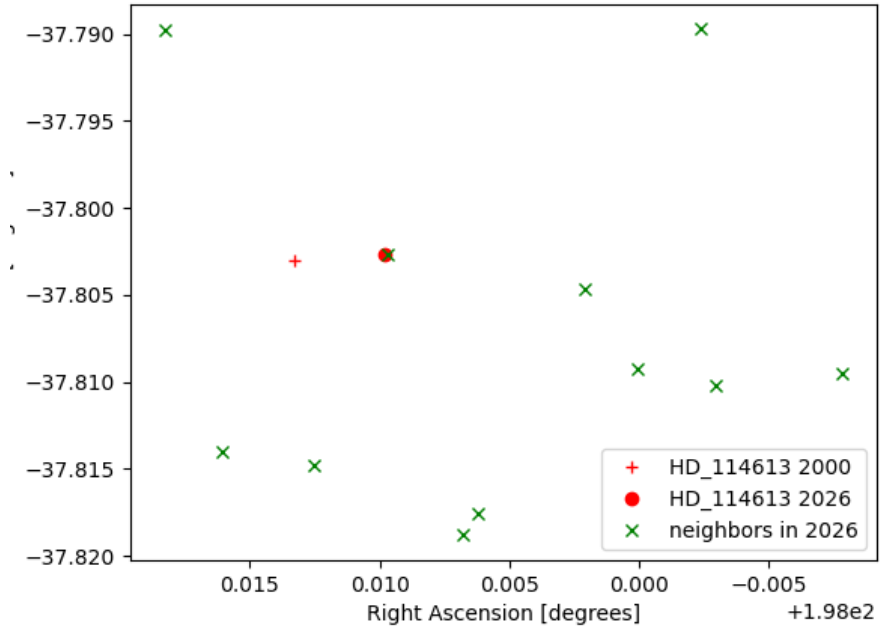

*Figure 28 Positions of HD 114613 in 2000 and 2026 compared with positions of other Gaia sources within one arcminute.*

*Table 15 Positions of stars within one arcminute of HD 114613 in 2026. Target star in bold italics.*

| Source ID | Distance (arcseconds) | RA | Dec | G mag |
|---|---|---|---|---|
| 6141740737212770176 | 58.6572 | 13h12m01.6255s | -37d49m07.7208s | 18.1022 |
| 6141740737211402880 | 54.4518 | 13h12m01.4888s | -37d49m03.1832s | 20.1172 |
| 6141740805930881408 | 45.18 | 13h11m59.2918s | -37d48m36.7003s | 20.1659 |
| 6141740801637679616 | 55.9515 | 13h11m58.1115s | -37d48m34.3893s | 18.0182 |
| 6141740805932250624 | 36.3211 | 13h12m00.0175s | -37d48m33.2825s | 16.9388 |
| 6141742214681522176 | 44.5623 | 13h12m03.8516s | -37d48m50.5352s | 16.6065 |
| ***6141742283401003264*** | ***0.21366*** | ***13h12m02.33s*** | ***-37d48m09.6579s*** | ***4.6041*** |
| 6141742317759093120 | 52.5585 | 13h12m04.3927s | -37d47m23.0554s | 19.7557 |
| 6141742214679865344 | 44.2042 | 13h12m03.0005s | -37d48m53.2137s | 16.741 |
| 6141742283399941120 | 23.1168 | 13h12m00.4953s | -37d48m16.9261s | 18.08 |
| 6141742489557784448 | 58.1325 | 13h11m59.4242s | -37d47m23.0169s | 19.8122 |

A.8 HD 134987

There is no Hubble data for HD 134987, so the analysis only includes the Gaia catalog and Besançon models. The first plot shows the comparison of the Gaia catalog to the Besançon models, showing source counts per square arcminute broken out by Gaia G magnitudes. The second figure and the table show the star positions from Gaia within one arcminute of the target star in 2026.

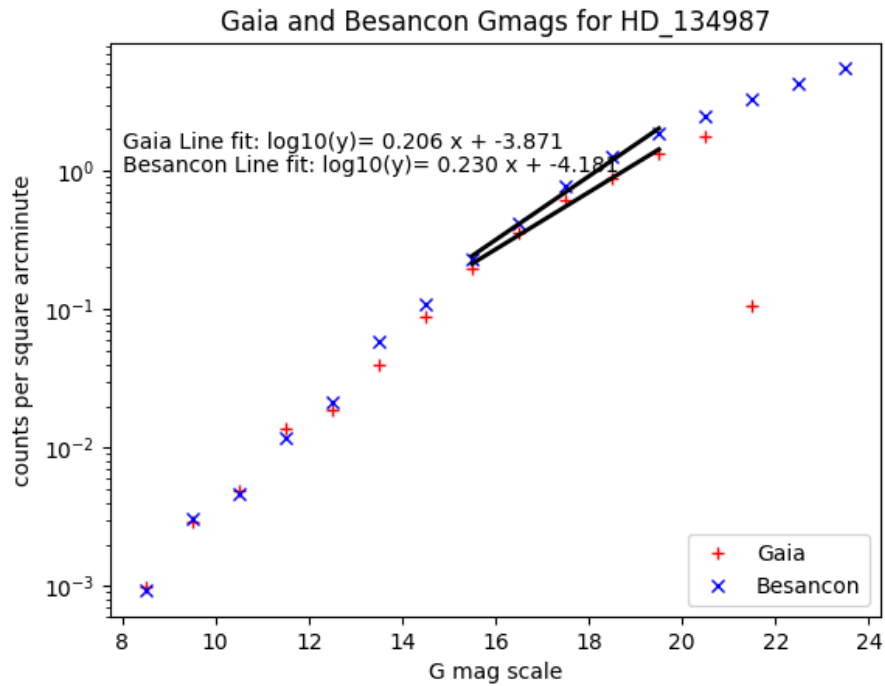

*Figure 29 Gaia counts per magnitude bin per square arcminute compared to Besançon simulations of the same for HD 134987.*

The position of HD 134987 in 2026 in RA, Dec is 15h13m27.8995s -25d18m35.6008s as calculated from coordinates and proper motions found in SIMBAD in J2000 coordinates.

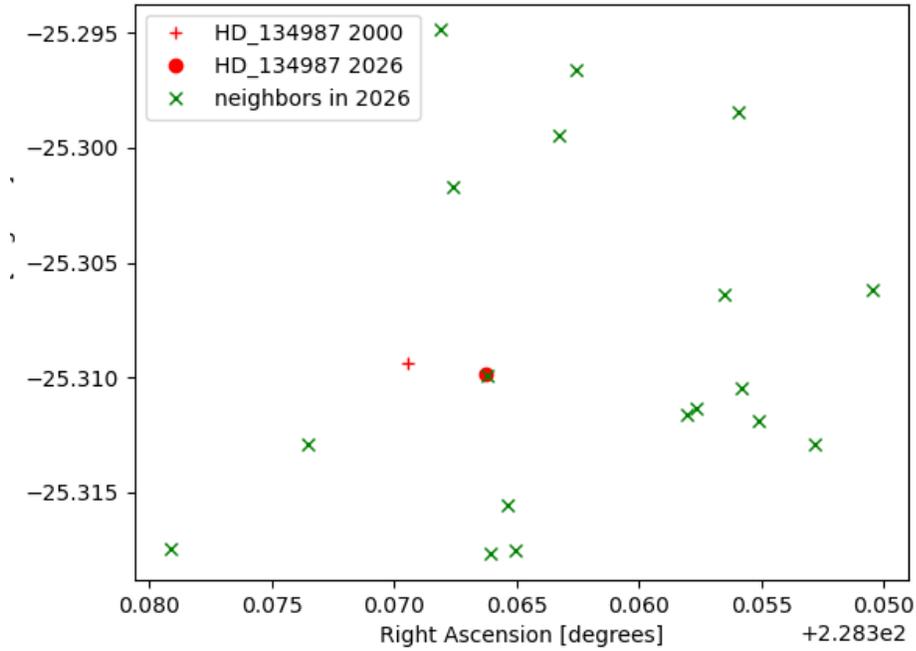

*Figure 30 Positions of HD 134987 in 2000 and 2026 compared with positions of other Gaia sources within one arcminute.*

*Table 16 Positions of stars within one arcminute of HD 134987 in 2026. Target star in bold italics.*

| Source ID | Distance (arcseconds) | RA | Dec | G mag |
|---|---|---|---|---|
| 6226694571852593792 | 27.727 | 15h13m27.6182s | -25d19m03.0643s | 20.623 |
| 6226694571852594304 | 28.1001 | 15h13m27.8595s | -25d19m03.6957s | 20.6672 |
| 6226694571852171648 | 36.8607 | 15h13m25.2339s | -25d18m42.8257s | 20.2321 |
| 6226694571852171264 | 28.4234 | 15h13m25.8386s | -25d18m40.7861s | 18.9553 |
| ***6226694571859532800*** | ***0.20438*** | ***15h13m27.8848s*** | ***-25d18m35.6464s*** | ***6.2855*** |
| 6226694567557774976 | 20.6197 | 15h13m27.6893s | -25d18m56.0225s | 17.8468 |
| 6226694571852171904 | 33.9348 | 15h13m25.4022s | -25d18m37.789s | 19.1932 |
| 6226694571852171392 | 27.3736 | 15h13m25.9331s | -25d18m41.7891s | 19.7601 |
| 6226694571852171520 | 44.9868 | 15h13m24.6823s | -25d18m46.5853s | 19.7234 |
| 6226694773715602688 | 53.1791 | 15h13m24.1043s | -25d18m22.2082s | 18.6414 |
| 6226694778010603776 | 33.9635 | 15h13m25.5704s | -25d18m23.1117s | 18.9884 |
| 6226694773715604352 | 53.1817 | 15h13m25.4247s | -25d17m54.3455s | 19.4535 |
| 6226695980608801664 | 50.0686 | 15h13m30.9983s | -25d19m02.8284s | 18.8917 |
| 6226696045026524928 | 29.8436 | 15h13m28.2188s | -25d18m06.073s | 19.4181 |
| 6226695980601424640 | 26.0881 | 15h13m29.6512s | -25d18m46.3885s | 20.6085 |
| 6226696255486715648 | 49.3556 | 15h13m27.023s | -25d17m47.6979s | 17.8724 |

| 6226696255486827520 | 38.7852 | 15h13m27.1902s | -25d17m58.0274s | 17.2242 |
| 6226696255479356288 | 54.4068 | 15h13m28.3505s | -25d17m41.5389s | 19.3674 |

A.9 HD 154345

There is no Hubble data for HD 154345, so the analysis only includes the Gaia catalog and Besançon models. The first plot shows the comparison of the Gaia catalog to the Besançon models, showing source counts per square arcminute broken out by Gaia G magnitudes. The second figure and the table show the star positions from Gaia within one arcminute of the target star in 2026.

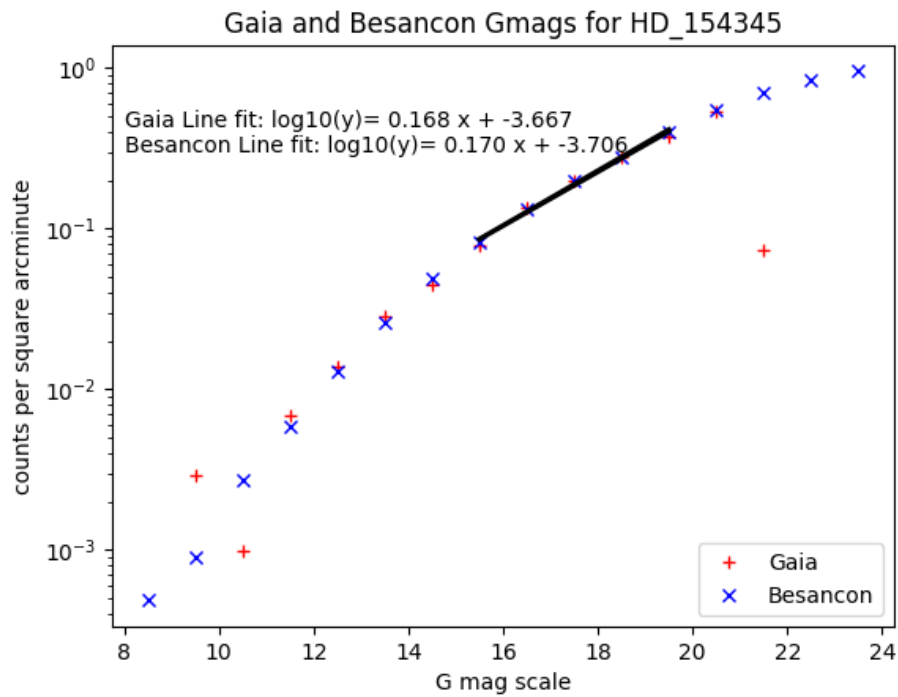

*Figure 31 Gaia counts per magnitude bin per square arcminute compared to Besançon simulations of the same for HD 154345.*

The position of HD 154345 in 2026 in RA, Dec is 17h02m36.7172s +47d05m16.9614s as calculated from coordinates and proper motions found in SIMBAD in J2000 coordinates.

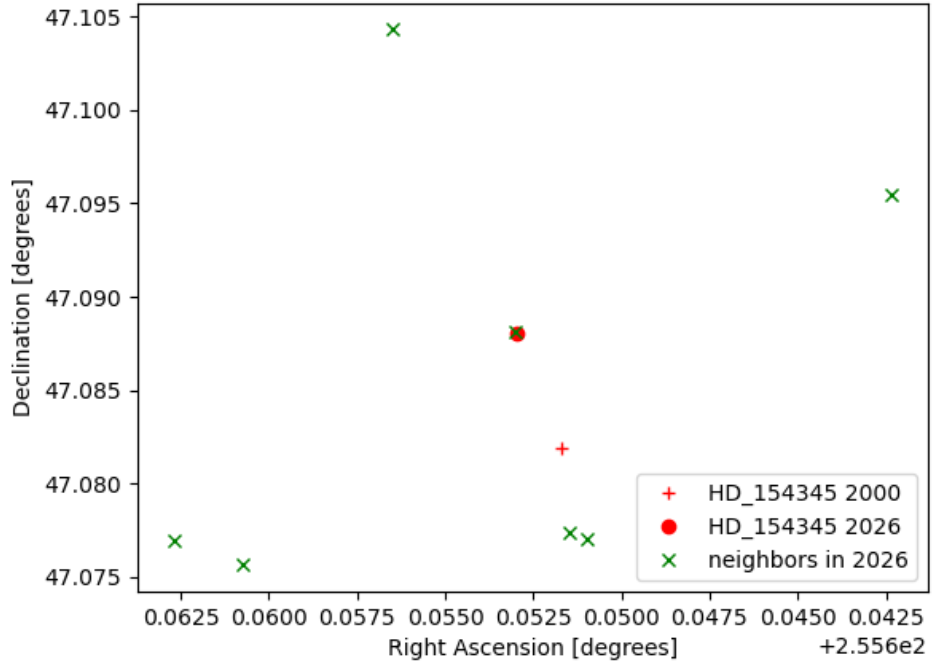

*Figure 32 Positions of HD 154345 in 2000 and 2026 compared with positions of other Gaia sources within one arcminute.*

*Table 17 Positions of stars within one arcminute of HD 154345 in 2026. Target star in bold italics.*

| Source ID | Distance (arcseconds) | RA | Dec | G mag |
|---|---|---|---|---|
| 1359938348454023936 | 46.4644 | 17h02m39.0496s | +47d04m37.069s | 20.357 |
| 1359938348454024064 | 48.4967 | 17h02m38.574s | +47d04m32.3269s | 18.995 |
| 1359938520252717056 | 39.8187 | 17h02m36.2322s | +47d04m37.452s | 18.0025 |
| ***1359938520253565952*** | ***0.43034*** | ***17h02m36.7236s*** | ***+47d05m17.3868s*** | ***6.5677*** |
| 1359938520253565184 | 38.5834 | 17h02m36.3569s | +47d04m38.554s | 16.5179 |
| 1359939344886446464 | 37.4041 | 17h02m34.1579s | +47d05m43.7176s | 19.7019 |
| 1359939374952210048 | 59.132 | 17h02m37.5626s | +47d06m15.4599s | 20.8453 |

A.10 HD 160691

There were no Besançon simulations run for this target, so the comparison plot for Gaia and Besançon do not exist. There is no Hubble data for HD 160691, so the analysis only includes the Gaia catalog. The figure and the table show the star positions from Gaia within one arcminute of the target star in 2026.

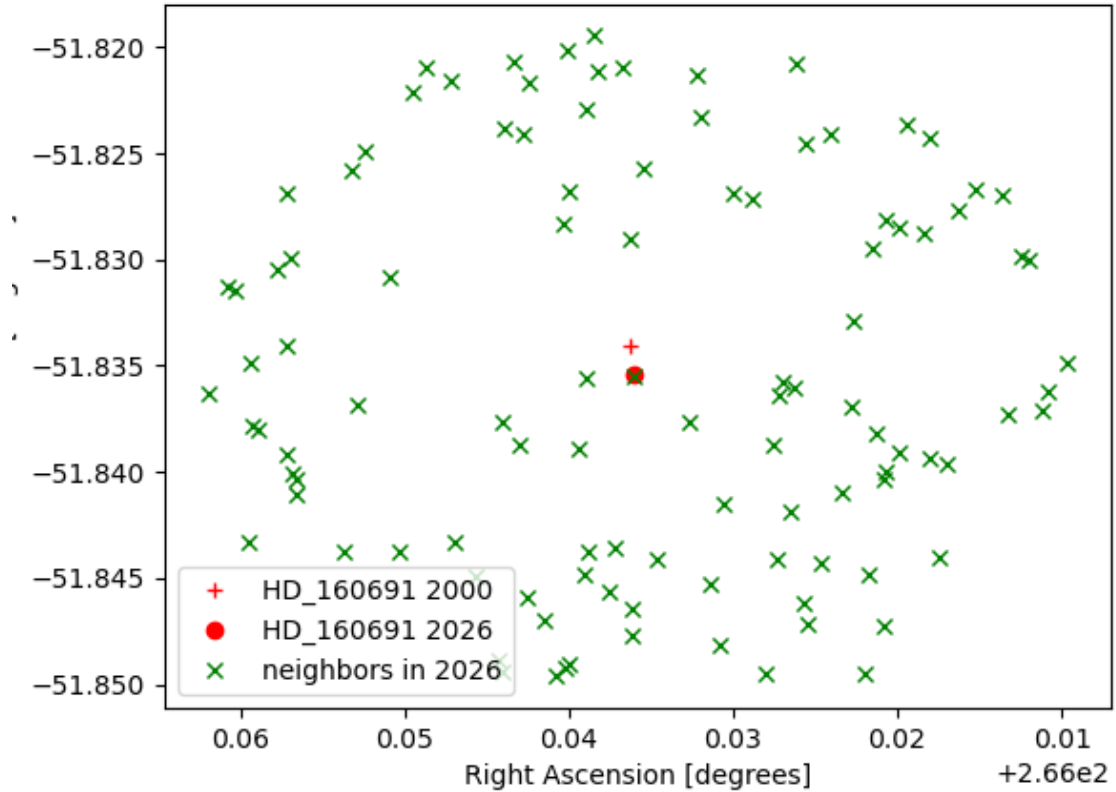

Figure 33 Positions of HD 160691 in 2000 and 2026 compared with positions of other Gaia sources within one arcminute.

Table 18 Positions of stars within one arcminute of HD 160691 in 2026. Target star in bold italics.

| Source ID | Distance (arcseconds) | RA | Dec | G mag |
|---|---|---|---|---|
| ***5945941905576552064*** | ***0.09581*** | ***17h44m08.6599s*** | ***-51d50m07.6504s*** | ***4.9046*** |
| 5945941905576873472 | 6.31182 | 17h44m09.3401s | -51d50m07.9785s | 14.9915 |
| 5945941905569456512 | 10.9238 | 17h44m07.8435s | -51d50m15.4264s | 16.4316 |
| 5946035742006792064 | 23.0674 | 17h44m08.6924s | -51d49m44.4894s | 17.9404 |
| 5945941905562631680 | 14.4635 | 17h44m09.4609s | -51d50m19.9717s | 14.8977 |
| 5945941905569222656 | 19.2842 | 17h44m10.5566s | -51d50m15.4975s | 18.1123 |
| 5945941905569228288 | 19.9302 | 17h44m06.5403s | -51d50m10.8681s | 16.6731 |
| 5945941905576800256 | 20.2583 | 17h44m06.4793s | -51d50m08.8135s | 14.9379 |
| 5945941905569222144 | 19.2942 | 17h44m10.3128s | -51d50m19.2934s | 17.67 |
| 5946035737715783296 | 27.3942 | 17h44m09.6878s | -51d49m41.8683s | 16.6439 |
| 5945941905562677632 | 21.9244 | 17h44m06.3071s | -51d50m09.7477s | 18.7326 |
| 5945941901271338880 | 22.4174 | 17h44m06.6164s | -51d50m19.5355s | 14.5173 |
| 5946035742023713280 | 32.4071 | 17h44m09.5849s | -51d49m36.3008s | 18.7669 |
| 5946035742023275392 | 33.6631 | 17h44m07.1914s | -51d49m36.7703s | 17.859 |
| 5945941871209484544 | 24.9812 | 17h44m07.3331s | -51d50m29.2959s | 18.4466 |
| 5946035742023275648 | 33.9983 | 17h44m06.9216s | -51d49m37.6213s | 15.596 |
| 5946035742024159872 | 35.0265 | 17h44m08.5029s | -51d49m32.559s | 19.6234 |
| 5945941905569231104 | 31.1399 | 17h44m05.4505s | -51d49m58.3723s | 18.8706 |
| 5945941901271338368 | 30.0793 | 17h44m05.4687s | -51d50m12.9815s | 17.0474 |
| 5945941905562619520 | 29.4287 | 17h44m08.9152s | -51d50m36.8889s | 17.7787 |
| 5946023990992752512 | 36.8566 | 17h44m12.205s | -51d49m50.8503s | 15.9747 |
| 5945941901271339136 | 30.6157 | 17h44m09.3048s | -51d50m37.5829s | 17.3596 |
| 5945941871217025024 | 31.5034 | 17h44m06.3734s | -51d50m30.8589s | 17.249 |
| 5945941871209483136 | 31.3753 | 17h44m08.3015s | -51d50m38.7531s | 17.6581 |
| 5945941871209750016 | 34.4697 | 17h44m05.1016s | -51d50m17.5544s | 19.2797 |
| 5945941974296277120 | 38.9538 | 17h44m05.1537s | -51d49m46.0919s | 16.8212 |
| 5945941871209486848 | 34.592 | 17h44m05.6224s | -51d50m27.6455s | 18.3791 |
| 5946023990974817792 | 37.7416 | 17h44m12.6976s | -51d50m12.4941s | 18.7399 |
| 5946035742005343104 | 43.4967 | 17h44m10.2581s | -51d49m26.6565s | 18.324 |
| 5945941871209716864 | 34.6352 | 17h44m09.372s | -51d50m41.5568s | 19.8187 |
| 5946035742005344896 | 44.5732 | 17h44m07.6806s | -51d49m23.9174s | 18.3769 |
| 5946035742005342976 | 45.1591 | 17h44m10.5413s | -51d49m25.8963s | 19.7713 |
| 5946035742008662784 | 45.5533 | 17h44m09.3333s | -51d49m22.4304s | 20.6462 |
| 5945941974282190592 | 43.0427 | 17h44m04.973s | -51d49m41.3978s | 18.8708 |
| 5945941935631076480 | 38.0497 | 17h44m04.9545s | -51d50m23.9198s | 17.479 |
| 5945941939936502656 | 38.4836 | 17h44m04.765s | -51d50m20.8683s | 19.0567 |
| 5945941901271339776 | 37.2799 | 17h44m11.2755s | -51d50m35.8835s | 19.0664 |
| 5945941939936292608 | 38.3411 | 17h44m04.9847s | -51d50m25.1403s | 18.799 |
| 5945941871216811776 | 36.7634 | 17h44m06.5667s | -51d50m38.7786s | 17.9894 |
| 5945941974282191872 | 43.77 | 17h44m04.7759s | -51d49m42.6739s | 19.1204 |
| 5945941871209483264 | 36.8252 | 17h44m07.5382s | -51d50m42.8802s | 18.7271 |
| 5945941871216806016 | 36.9105 | 17h44m09.0042s | -51d50m44.3278s | 17.3887 |
| 5946035742008300672 | 45.6879 | 17h44m06.1411s | -51d49m28.2881s | 20.5369 |
| 5945941974289607936 | 46.0624 | 17h44m04.4246s | -51d49m43.4744s | 20.6961 |
| 5945941905576541824 | 40.2302 | 17h44m10.9728s | -51d50m41.6032s | 14.6803 |
| 5945941871209481984 | 39.5316 | 17h44m08.6718s | -51d50m47.0863s | 19.1092 |
| 5945941939922425856 | 42.5437 | 17h44m04.328s | -51d50m21.6046s | 19.4973 |
| 5945941905562598016 | 40.3638 | 17h44m10.2012s | -51d50m45.3095s | 19.6427 |
| 5945941866911600128 | 40.7952 | 17h44m05.9176s | -51d50m39.4596s | 18.4401 |
| 5946035742005346048 | 48.8398 | 17h44m05.778s | -51d49m26.674s | 19.1097 |
| 5946023990992752000 | 47.017 | 17h44m13.7064s | -51d50m02.7282s | 19.0851 |
| 5946035742005859072 | 51.4612 | 17h44m07.7195s | -51d49m16.8387s | 20.1476 |
| 5946035742005343616 | 51.553 | 17h44m10.1871s | -51d49m17.982s | 19.2856 |
| 5945930154546276736 | 43.5193 | 17h44m12.0781s | -51d50m37.4004s | 20.0763 |
| 5946035742006801792 | 51.7665 | 17h44m09.182s | -51d49m16.0145s | 20.2998 |
| 5946035742005344768 | 52.0439 | 17h44m08.8018s | -51d49m15.5274s | 19.7463 |
| 5945941939936693376 | 44.9614 | 17h44m04.0875s | -51d50m22.5544s | 20.6963 |
| 5945941866911601024 | 43.2268 | 17h44m09.9542s | -51d50m49.0861s | 17.2813 |
| 5946024025337882368 | 51.4907 | 17h44m12.7787s | -51d49m32.9997s | 20.689 |
| 5946023986685261952 | 50.5628 | 17h44m13.6774s | -51d49m47.7019s | 18.0676 |
| 5946024025335058816 | 52.481 | 17h44m12.5839s | -51d49m29.7178s | 19.9848 |
| 5945941871216805120 | 44.1589 | 17h44m08.6881s | -51d50m51.7131s | 13.5659 |
| 5946023990974818688 | 51.3546 | 17h44m13.8514s | -51d49m49.602s | 18.8057 |
| 5945941866911600256 | 45.1779 | 17h44m06.173s | -51d50m46.4072s | 18.0819 |
| 5945941974287131904 | 52.0278 | 17h44m03.9141s | -51d49m39.7883s | 19.767 |
| 5946023990978150656 | 48.8646 | 17h44m13.7326s | -51d50m20.8948s | 20.5968 |
| 5946023990974742784 | 48.9814 | 17h44m13.6248s | -51d50m24.3567s | 19.9014 |
| 5945941871217138944 | 46.5976 | 17h44m05.2164s | -51d50m41.5019s | 20.9736 |
| 5945930154529261056 | 49.008 | 17h44m13.5869s | -51d50m25.3631s | 20.263 |
| 5946035776366538112 | 55.5456 | 17h44m10.3941s | -51d49m14.3844s | 20.3668 |
| 5946035742005344384 | 55.8005 | 17h44m09.6096s | -51d49m12.4521s | 19.0764 |
| 5946035776365587328 | 55.7045 | 17h44m11.3138s | -51d49m17.574s | 20.3248 |
| 5946023990974816896 | 51.8054 | 17h44m14.2462s | -51d50m05.6831s | 18.6523 |
| 5945930154542359424 | 49.8665 | 17h44m13.5716s | -51d50m27.9268s | 21.0499 |
| 5945930154538696704 | 49.2614 | 17h44m12.8851s | -51d50m37.4518s | 19.8596 |
| 5945941871217101184 | 47.3292 | 17h44m07.3946s | -51d50m53.4066s | 20.7406 |
| 5946035776365081088 | 56.5514 | 17h44m11.8899s | -51d49m19.5764s | 18.8032 |
| 5946023990974816384 | 51.637 | 17h44m14.1402s | -51d50m16.8861s | 19.9265 |
| 5945941939928968832 | 51.2533 | 17h44m03.1797s | -51d50m14.3463s | 20.1949 |
| 5946035742008301312 | 57.7196 | 17h44m09.2277s | -51d49m10.0751s | 20.8396 |
| 5946035737715782016 | 57.1461 | 17h44m06.2751s | -51d49m14.8616s | 16.8913 |
| 5945941974288935296 | 56.2711 | 17h44m04.6524s | -51d49m25.2964s | 20.2383 |
| 5945941866907658112 | 48.5223 | 17h44m06.0986s | -51d50m49.8707s | 18.3937 |
| 5946023990974816512 | 52.3147 | 17h44m14.2279s | -51d50m16.1721s | 19.199 |
| 5946024025334557824 | 56.0899 | 17h44m13.7285s | -51d49m36.9076s | 19.6317 |
| 5945941974292362880 | 55.8955 | 17h44m03.6664s | -51d49m36.2343s | 20.7011 |
| 5945941974288712320 | 56.7775 | 17h44m04.335s | -51d49m27.3584s | 19.9265 |
| 5946023990992752256 | 55.7465 | 17h44m14.4721s | -51d49m53.2023s | 17.9279 |
| 5945941871202852608 | 49.7668 | 17h44m09.5977s | -51d50m56.5582s | 20.3354 |
| 5946035776365081216 | 59.2196 | 17h44m11.6829s | -51d49m15.3819s | 19.0692 |
| 5945941974282209664 | 56.3559 | 17h44m02.9857s | -51d49m47.3344s | 20.8716 |
| 5945941939936537344 | 51.7512 | 17h44m04.197s | -51d50m38.6481s | 20.2121 |
| 5945941871202852736 | 50.4714 | 17h44m09.6456s | -51d50m57.1941s | 18.5215 |
| 5946023990974817920 | 56.976 | 17h44m14.5935s | -51d49m52.6548s | 19.921 |
| 5945941974282209792 | 57.0175 | 17h44m02.8823s | -51d49m48.0076s | 19.4362 |
| 5945941871209716352 | 51.7623 | 17h44m10.6152s | -51d50m56.0447s | 20.3619 |
| 5945941974296041472 | 58.5687 | 17h44m03.2598s | -51d49m37.1611s | 19.6096 |
| 5945941871216802048 | 52.1627 | 17h44m09.787s | -51d50m58.6626s | 15.6888 |
| 5945941935627136640 | 55.6993 | 17h44m02.6872s | -51d50m13.6352s | 16.1987 |
| 5945941939930770816 | 56.1463 | 17h44m02.6106s | -51d50m10.3407s | 20.711 |
| 5946023990974816256 | 57.6954 | 17h44m14.8755s | -51d50m10.8079s | 19.5768 |
| 5945941871202841344 | 53.3949 | 17h44m10.5629s | -51d50m57.9558s | 19.8124 |
| 5945941871202884992 | 53.6364 | 17h44m06.7185s | -51d50m58.0811s | 19.5996 |
| 5945941871202913280 | 54.3182 | 17h44m05.0029s | -51d50m49.997s | 19.8621 |
| 5945941939928970880 | 58.9187 | 17h44m02.3077s | -51d50m05.5736s | 20.1843 |
| 5945930154532040064 | 59.228 | 17h44m14.2764s | -51d50m35.8228s | 18.76 |
| 5945941768130267776 | 59.7688 | 17h44m05.261s | -51d50m58.3455s | 20.0679 |

A.11 HD 190360

There is no Hubble data for HD 190360, so the analysis only includes the Gaia catalog and Besançon models. The first plot shows the comparison of the Gaia catalog to the Besançon models, showing source counts per square arcminute broken out by Gaia G magnitudes. The second figure and the table show the star positions from Gaia within one arcminute of the target star in 2026.

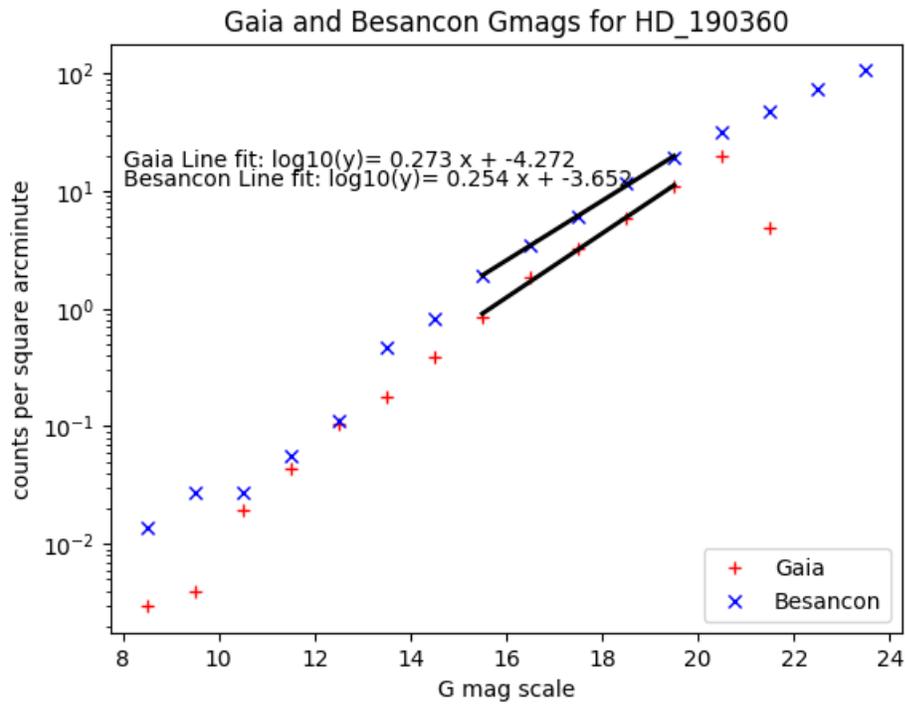

*Figure 34 Gaia counts per magnitude bin per square arcminute compared to Besançon simulations of the same for HD 190360.*

The position of HD 190360 in 2026 in RA, Dec is 20h03m38.7721s +29d53m34.8264s as calculated from coordinates and proper motions found in SIMBAD in J2000 coordinates.

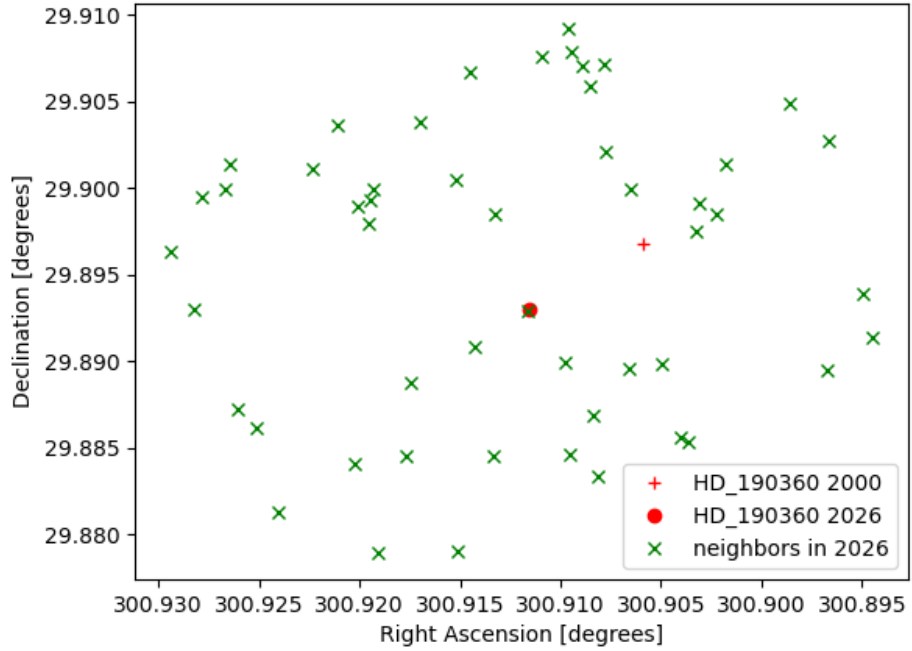

*Figure 35 Positions of HD 190360 in 2000 and 2026 compared with positions of other Gaia sources within one arcminute.*

*Table 19 Positions of stars within one arcminute of HD 190360 in 2026. Target star in bold italics.*

| Source ID | Distance (arcseconds) | RA | Dec | G mag |
|---|---|---|---|---|
| 2029432765329761664 | 31.1654 | 20h03m39.2026s | +29d53m04.1681s | 20.764 |
| 2029432765330391040 | 11.6278 | 20h03m39.4284s | +29d53m26.9295s | 19.1538 |
| 2029432765327143552 | 36.0466 | 20h03m40.2361s | +29d53m04.2183s | 20.4234 |
| 2029432765328343040 | 30.8263 | 20h03m38.2952s | +29d53m04.6303s | 20.4926 |
| 2029432761021700096 | 55.996 | 20h03m40.577s | +29d52m43.988s | 19.6152 |
| 2029432761021271040 | 51.5863 | 20h03m39.6298s | +29d52m44.4607s | 18.7672 |
| 2029432761025276032 | 23.9488 | 20h03m40.1884s | +29d53m19.5201s | 19.2462 |
| 2029432761025277056 | 42.1644 | 20h03m40.8568s | +29d53m02.5344s | 20.5324 |
| 2029432761025278464 | 57.3239 | 20h03m41.7632s | +29d52m52.7214s | 20.6072 |
| 2029432765329890560 | 36.2575 | 20h03m37.9518s | +29d53m00.1738s | 21.1459 |
| 2029432868406359680 | 53.5394 | 20h03m42.4109s | +29d53m59.8757s | 19.9149 |
| 2029432864100921856 | 55.359 | 20h03m42.3394s | +29d54m05.0376s | 19.7199 |
| 2029432868407576320 | 56.0215 | 20h03m42.6835s | +29d53m58.3072s | 20.3794 |
| 2029432864104494080 | 49.0618 | 20h03m42.0374s | +29d53m10.2548s | 20.1551 |
| 2029432864100489728 | 49.7828 | 20h03m42.2535s | +29d53m14.1252s | 16.5718 |
| 2029432864100493056 | 52.157 | 20h03m42.7828s | +29d53m34.919s | 17.7616 |
| 2029432868405734912 | 57.0936 | 20h03m43.0627s | +29d53m46.9298s | 18.0439 |
| 2029433521244006784 | 30.569 | 20h03m36.7724s | +29d53m50.8958s | 20.8184 |
| 2029433516939518208 | 19.938 | 20h03m37.5757s | +29d53m22.358s | 19.8272 |
| 2029433589962082176 | 42.7916 | 20h03m36.4274s | +29d54m04.8513s | 20.4503 |
| 2029433521244691968 | 34.2551 | 20h03m36.7506s | +29d53m56.7889s | 19.571 |
| 2029433585659030144 | 58.2236 | 20h03m35.2039s | +29d54m09.9973s | 20.3816 |
| 2029433585659030656 | 58.7987 | 20h03m35.6611s | +29d54m17.4966s | 19.5716 |
| 2029433486889134976 | 53.7004 | 20h03m34.6669s | +29d53m29.0166s | 19.8075 |
| 2029433521244261376 | 23.6316 | 20h03m37.1866s | +29d53m23.2797s | 20.1811 |
| 2029433521248542464 | 24.2345 | 20h03m38.0099s | +29d53m12.7116s | 17.013 |
| *2029433521248546304* | *0.42062* | *20h03m38.7974s* | *+29d53m34.565s* | *5.5336* |
| 2029433521248551424 | 29.4112 | 20h03m37.5673s | +29d53m59.7178s | 13.7435 |
| 2029433521240806656 | 35.511 | 20h03m36.9734s | +29d53m08.1078s | 21.0218 |
| 2029433585658992128 | 52.0102 | 20h03m34.78s | +29d53m37.9795s | 20.4412 |
| 2029433516939515776 | 47.8681 | 20h03m35.2204s | +29d53m22.2559s | 18.8095 |
| 2029433516939517440 | 37.0356 | 20h03m36.8757s | +29d53m07.197s | 18.5789 |
| 2029433516939518720 | 12.3187 | 20h03m38.3471s | +29d53m23.8173s | 14.0642 |
| 2029433521243180928 | 35.1966 | 20h03m36.5331s | +29d53m54.6008s | 19.7191 |
| 2029433624320003712 | 33.5665 | 20h03m40.6793s | +29d53m57.4452s | 19.1763 |
| 2029433693040092544 | 52.1705 | 20h03m37.8727s | +29d54m25.6692s | 20.1795 |
| 2029433624322377344 | 29.1519 | 20h03m39.6449s | +29d54m01.6778s | 20.5879 |
| 2029433693040092032 | 54.0621 | 20h03m38.2603s | +29d54m28.4774s | 18.9553 |
| 2029433624321823232 | 50.205 | 20h03m39.479s | +29d54m24.1826s | 19.1294 |
| 2029433620018771200 | 47.4362 | 20h03m38.0486s | +29d54m21.3202s | 20.2029 |
| 2029433688738248576 | 58.5633 | 20h03m38.3154s | +29d54m33.0878s | 18.4581 |
| 2029433620018771968 | 52.5572 | 20h03m38.6247s | +29d54m27.3486s | 18.3042 |
| 2029433620018774656 | 48.5126 | 20h03m41.0657s | +29d54m13.0877s | 20.9952 |
| 2029433624327755776 | 34.7516 | 20h03m40.6247s | +29d53m59.8719s | 17.459 |
| 2029433624322965888 | 30.6891 | 20h03m40.6984s | +29d53m52.5558s | 20.5654 |
| 2029433624322373888 | 42.5533 | 20h03m40.0801s | +29d54m13.8324s | 20.6814 |
| 2029433624327767552 | 34.8674 | 20h03m37.8615s | +29d54m07.6214s | 17.9071 |
| 2029433624327769344 | 51.1189 | 20h03m38.1403s | +29d54m25.281s | 18.3089 |
| 2029433620018734464 | 20.618 | 20h03m39.1876s | +29d53m54.724s | 20.3351 |
| 2029433620018736000 | 34.2073 | 20h03m40.8175s | +29d53m56.3362s | 18.7808 |
| 2029433624327915264 | 44.6335 | 20h03m41.3593s | +29d54m04.1561s | 19.0579 |

A.12 HD 192310

There is no Hubble data for HD 192310, so the analysis only includes the Gaia catalog and Besançon models. The first plot shows the comparison of the Gaia catalog to the Besançon models, showing source counts per square arcminute broken out by Gaia G magnitudes. The second figure and the table show the star positions from Gaia within one arcminute of the target star in 2026.

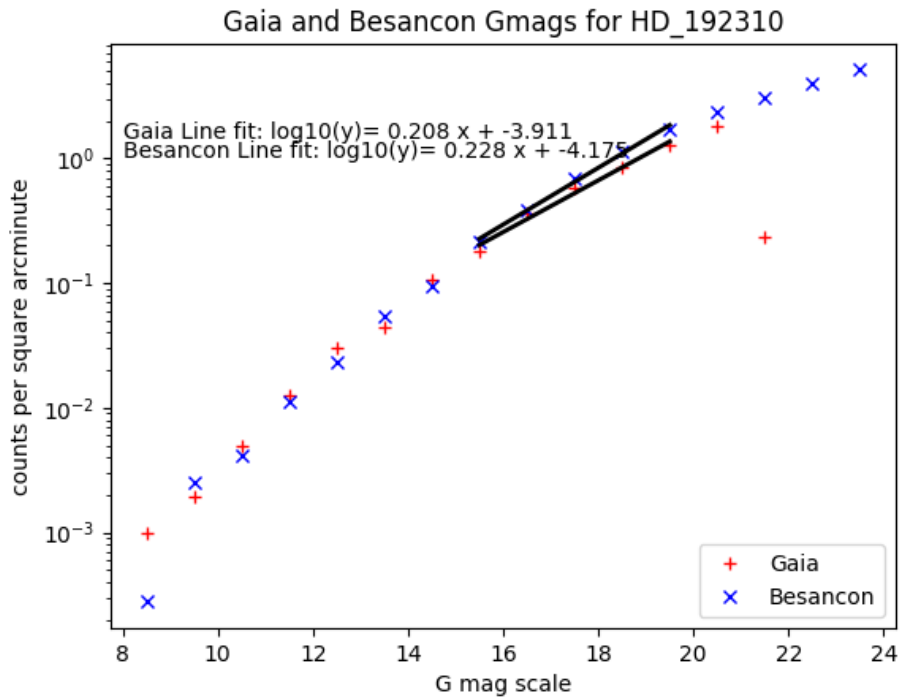

Figure 36 Gaia counts per magnitude bin per square arcminute compared to Besançon simulations of the same for HD 192310.

The position of HD 192310 in 2026 in RA, Dec is 20h15m19.8091s -27d02m03.418s as calculated from coordinates and proper motions found in SIMBAD in J2000 coordinates.

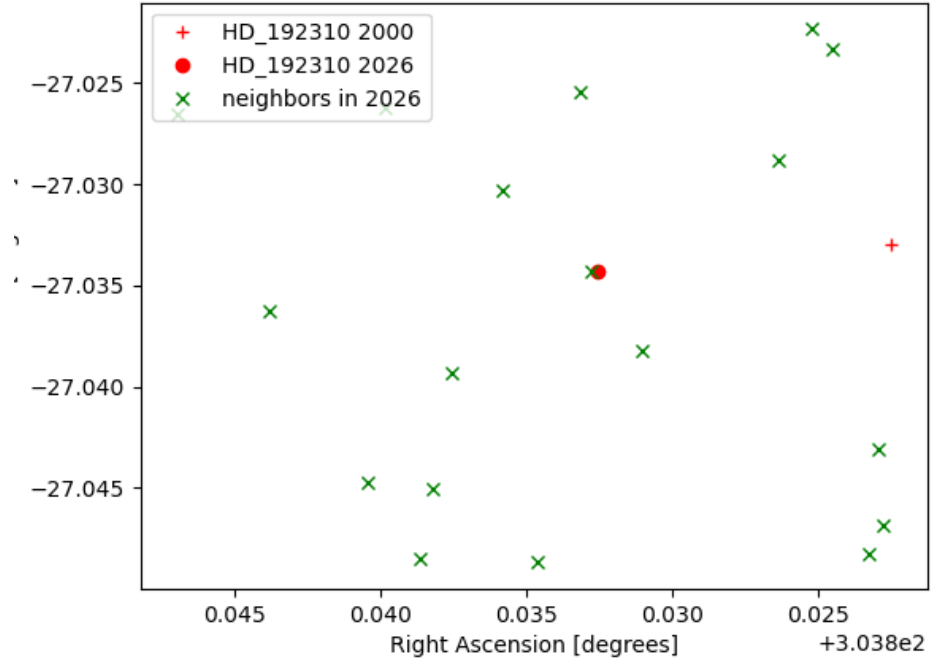

*Figure 37. Positions of HD 192310 in 2000 and 2026 compared with positions of other Gaia sources within one arcminute.*

*Table 20 Positions of stars within one arcminute of HD 192310 in 2026. Target star in bold italics.*

| Source ID | Distance (arcseconds) | RA | Dec | G mag |
|---|---|---|---|---|
| 6847164651447229824 | 14.8892 | 20h15m19.4504s | -27d02m17.5145s | 19.2364 |
| 6847164647150392448 | 36.6899 | 20h15m22.5026s | -27d02m10.5662s | 19.3436 |
| 6847164651446895488 | 17.693 | 20h15m20.5916s | -27d01m49.1442s | 18.4647 |
| 6847164617088100864 | 52.2473 | 20h15m20.3002s | -27d02m55.2518s | 16.2898 |
| 6847164647150384512 | 45.4111 | 20h15m21.7004s | -27d02m41.1497s | 19.269 |
| 6847164647150384384 | 42.8784 | 20h15m21.1715s | -27d02m42.2414s | 17.5829 |
| 6847164651447693952 | 24.0703 | 20h15m21.0043s | -27d02m21.4283s | 15.3184 |
| 6847164617085484672 | 58.3593 | 20h15m17.5844s | -27d02m53.6408s | 19.6933 |
| 6847164617085486976 | 44.2277 | 20h15m17.4997s | -27d02m35.1048s | 19.7001 |
| 6847164617085485312 | 55.1083 | 20h15m17.4624s | -27d02m48.7384s | 20.5997 |
| 6847164582725738624 | 54.7861 | 20h15m21.2752s | -27d02m54.5829s | 20.1428 |
| ***6847167606385195648*** | ***0.6338*** | ***20h15m19.856s*** | ***-27d02m03.5115s*** | ***5.4532*** |
| 6847167602089745280 | 49.0277 | 20h15m18.0482s | -27d01m20.4053s | 19.3449 |
| 6847167606384717952 | 31.9246 | 20h15m19.9541s | -27d01m31.5522s | 20.9071 |
| 6847167606384777472 | 28.0007 | 20h15m18.3239s | -27d01m43.663s | 20.6226 |
| 6847167606382733824 | 47.0421 | 20h15m17.8834s | -27d01m24.0365s | 18.9931 |
| 6847170522665663872 | 53.9561 | 20h15m23.2711s | -27d01m35.6403s | 17.9361 |
| 6847173481897986048 | 37.3216 | 20h15m21.5544s | -27d01m34.2786s | 20.4011 |

A.13 HD 217107

There is no Hubble data for HD 217107, so the analysis only includes the Gaia catalog and Besançon models. The first plot shows the comparison of the Gaia catalog to the Besançon models, showing source counts per square arcminute broken out by Gaia G magnitudes. The second figure and the table show the star positions from Gaia within one arcminute of the target star in 2026.

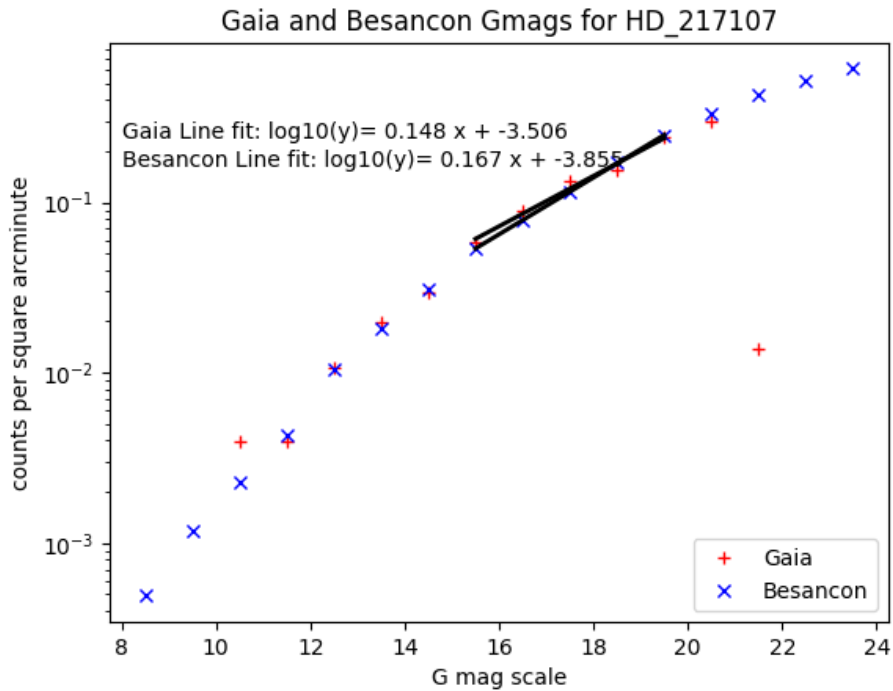

*Figure 38 Gaia counts per magnitude bin per square arcminute compared to Besançon simulations of the same for HD 217107.*

The position of HD 217107 in 2026 in RA, Dec is 22h58m15.5288s -02d23m43.7682s as calculated from coordinates and proper motions found in SIMBAD in J2000 coordinates.

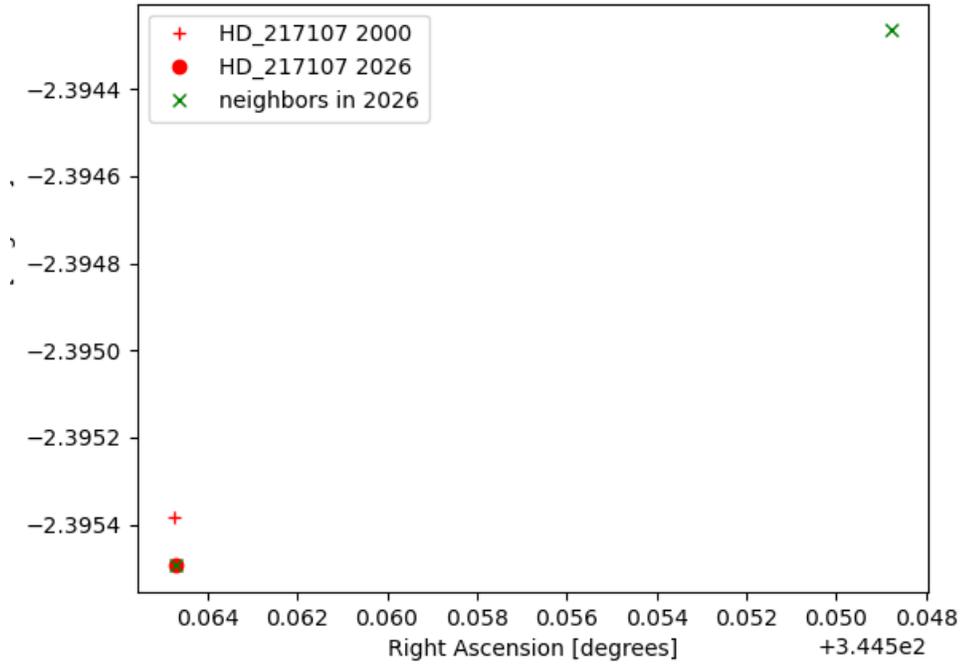

*Figure 39 Positions of HD 217107 in 2000 and 2026 compared with positions of other Gaia sources within one arcminute.*

*Table 21 Positions of stars within one arcminute of HD 217107 in 2026. Target star in bold italics.*

| Source ID | Distance (arcseconds) | RA | Dec | G mag |
|---|---|---|---|---|
| 2648914005997582848 | 57.5421 | 22h58m11.7006s | -02d23m39.3569s | 17.1701 |
| ***2648914040357320576*** | ***0.0109*** | ***22h58m15.5286s*** | ***-02d23m43.7787s*** | ***5.9546*** |

## A.14 HD 219134

There is no Hubble data for HD 219134, so the analysis only includes the Gaia catalog and Besançon models. The first plot shows the comparison of the Gaia catalog to the Besançon models, showing source counts per square arcminute broken out by Gaia G magnitudes. The second figure and the table show the star positions from Gaia within one arcminute of the target star in 2026.

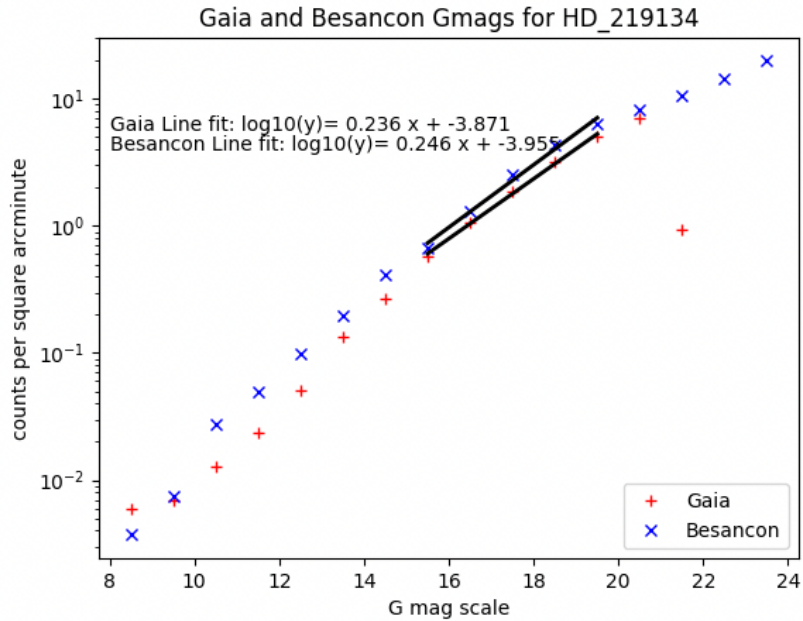

*Figure 40 Gaia counts per magnitude bin per square arcminute compared to Besançon simulations of the same for HD 219134.*

The position of HD 219134 in 2026 is 23h13m23.6074s +57d10m13.7339s as calculated from coordinates and proper motions found in SIMBAD in J2000 coordinates.

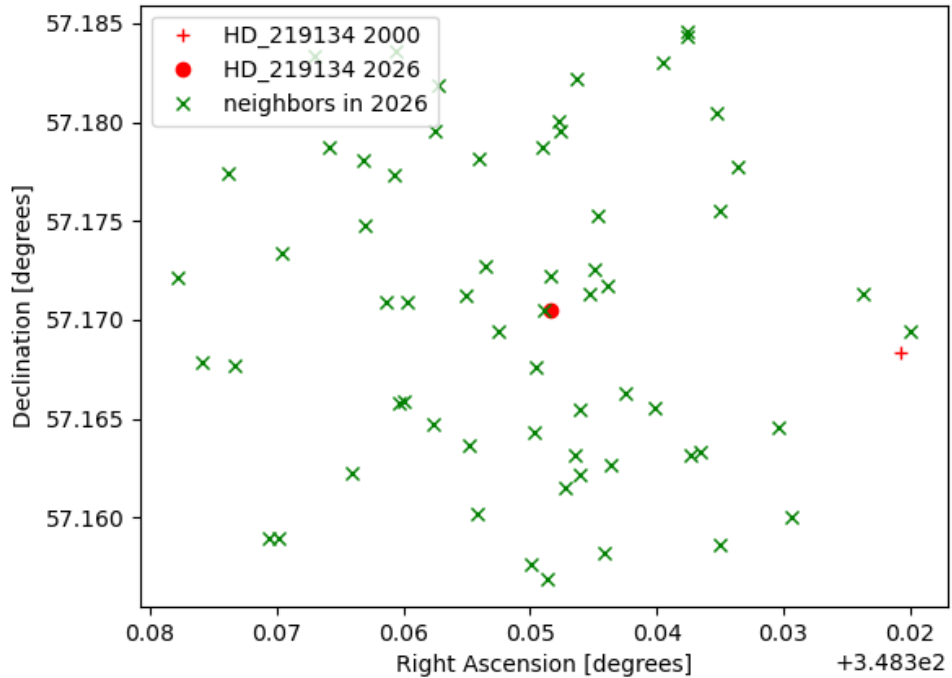

*Figure 41 Positions of HD 219134 in 2000 and 2026 compared with positions of other Gaia sources within one arcminute.*

*Table 22 Positions of stars within one arcminute of HD 219134 in 2026. Target star in bold italics.*

| Source ID | Distance (arcseconds) | RA | Dec | G mag |
|---|---|---|---|---|
| 2009481027319050496 | 55.5486 | 23h13m29.7097s | +57d10m38.7052s | 20.7285 |
| 2009480988666978944 | 34.2054 | 23h13m20.9459s | +57d09m47.2485s | 18.7406 |
| 2009481027319054208 | 57.8659 | 23h13m30.6863s | +57d10m19.629s | 20.2735 |
| 2009480988667022336 | 32.4368 | 23h13m23.318s | +57d09m41.3827s | 17.8235 |
| 2009481027319039616 | 22.0887 | 23h13m26.3162s | +57d10m15.371s | 20.9173 |
| 2009480992961023616 | 24.0147 | 23h13m21.6381s | +57d09m55.8395s | 18.7241 |
| 2009481027319042048 | 45.3103 | 23h13m27.8139s | +57d10m43.4516s | 20.759 |
| 2009480992960630912 | 10.664 | 23h13m23.8917s | +57d10m03.3235s | 20.5609 |
| 2009480958597538432 | 42.6651 | 23h13m27.3798s | +57d09m44.0858s | 20.2433 |
| 2009480992955886720 | 34.5459 | 23h13m20.7721s | +57d09m48.0109s | 19.0872 |
| 2009481023036052480 | 27.6722 | 23h13m25.8442s | +57d09m52.8815s | 19.0444 |
| 2009480954303554560 | 38.6277 | 23h13m24.9947s | +57d09m36.7908s | 18.4266 |
| 2009481023022883840 | 34.481 | 23h13m26.5718s | +57d10m38.3897s | 18.2558 |
| 2009481027315617792 | 13.3658 | 23h13m25.2191s | +57d10m16.3541s | 19.4456 |
| 2009481027317000576 | 25.337 | 23h13m26.7184s | +57d10m15.1225s | 20.3014 |
| 2009480992962046720 | 6.71491 | 23h13m22.8597s | +57d10m16.5826s | 20.1234 |
| 2009481027321303424 | 12.938 | 23h13m24.8504s | +57d10m21.8102s | 17.5092 |
| 2009480924242570496 | 46.271 | 23h13m23.9705s | +57d09m27.5573s | 18.8524 |
| 2009480992962047104 | 26.6005 | 23h13m23.1557s | +57d09m47.3884s | 17.2669 |
| 2009480992962046976 | 45.0629 | 23h13m22.5851s | +57d09m29.4448s | 17.232 |
| 2009480958596176256 | 59.0218 | 23h13m28.7722s | +57d09m32.2737s | 20.0757 |
| 2009480992961024256 | 41.1881 | 23h13m19.2794s | +57d09m52.3451s | 19.8544 |
| 2009481027315635200 | 28.0649 | 23h13m26.3919s | +57d09m57.1552s | 18.2504 |
| 2009480992955893760 | 49.9194 | 23h13m20.4083s | +57d09m31.1322s | 19.163 |
| 2009480992961569664 | 22.3878 | 23h13m23.9238s | +57d09m51.4945s | 20.3822 |
| 2009480992961569024 | 29.5696 | 23h13m22.4691s | +57d09m45.651s | 16.6586 |
| 2009481027321309568 | 27.6057 | 23h13m25.1619s | +57d09m49.1934s | 17.6908 |
| 2009481027321309312 | 28.7638 | 23h13m26.4747s | +57d09m56.8939s | 19.6326 |
| 2009480958596164992 | 54.5307 | 23h13m30.2103s | +57d10m04.2343s | 19.7227 |
| 2009480988672333440 | 52.9832 | 23h13m19.0423s | +57d09m35.9365s | 17.2657 |
| 2009480954312595584 | 49.691 | 23h13m29.5906s | +57d10m03.6542s | 19.5242 |
| 2009480958601837568 | 59.9895 | 23h13m28.9465s | +57d09m32.3454s | 19.6949 |
| 2009480992959292416 | 30.2805 | 23h13m23.0685s | +57d09m43.7724s | 19.9614 |
| 2009481027315609088 | 32.4989 | 23h13m27.126s | +57d10m29.1462s | 20.3882 |
| 2009481027319033600 | 8.91248 | 23h13m24.6048s | +57d10m10.0404s | 20.4815 |
| 2009481027321301888 | 39.8396 | 23h13m27.1601s | +57d10m41.1689s | 14.0575 |
| 2009480924239819008 | 49.0663 | 23h13m23.6796s | +57d09m24.6711s | 20.632 |
| 2009480988667027840 | 6.31675 | 23h13m23.6053s | +57d10m20.0506s | 14.693 |
| 2009480988667001344 | 19.0875 | 23h13m22.184s | +57d09m58.5573s | 17.5703 |
| 2009480992961914752 | 18.5491 | 23h13m23.0443s | +57d09m55.7591s | 17.3254 |
| 2009481027319048448 | 42.7215 | 23h13m28.7035s | +57d10m24.1189s | 20.5683 |
| 2009481778941007104 | 32.5727 | 23h13m23.4153s | +57d10m46.2691s | 17.3007 |
| 2009481783233285504 | 58.8928 | 23h13m28.0876s | +57d11m00.0095s | 20.7121 |
| 2009481783235542656 | 44.428 | 23h13m25.7296s | +57d10m54.6741s | 19.6712 |
| 2009481744577538688 | 18.6796 | 23h13m22.7291s | +57d10m30.9941s | 16.9988 |
| 2009481748876290560 | 10.1785 | 23h13m22.7661s | +57d10m21.27s | 20.2204 |
| 2009481748876290432 | 9.77799 | 23h13m22.5298s | +57d10m18.0723s | 18.7009 |
| 2009481748874726784 | 55.6533 | 23h13m16.7793s | +57d10m10.0056s | 19.5007 |
| 2009481783229836416 | 29.6363 | 23h13m23.7679s | +57d10m43.3414s | 19.7264 |
| 2009481783229834752 | 52.8477 | 23h13m26.5538s | +57d11m00.8399s | 20.2361 |
| 2009481783229839360 | 37.1854 | 23h13m25.8134s | +57d10m46.307s | 20.6047 |
| 2009481748870088704 | 38.7761 | 23h13m20.0743s | +57d10m39.7766s | 19.834 |
| 2009481783233276288 | 29.806 | 23h13m24.9578s | +57d10m41.4438s | 20.2626 |
| 2009481748870094208 | 48.2292 | 23h13m17.6873s | +57d10m16.6231s | 18.7532 |
| 2009481783229832448 | 34.4227 | 23h13m23.4597s | +57d10m48.1356s | 17.676 |
| *2009481748875806976* | *1.0478* | *23h13m23.7348s* | *+57d10m13.8942s* | *5.2079* |
| 2009481817589559936 | 48.3121 | 23h13m21.4863s | +57d10m58.8631s | 19.4796 |
| 2009481748870085376 | 44.1261 | 23h13m20.4651s | +57d10m49.7102s | 20.0844 |
| 2009481817595273600 | 54.8275 | 23h13m21.0225s | +57d11m04.3734s | 14.3557 |
| 2009481783235539840 | 42.221 | 23h13m23.13s | +57d10m55.7761s | 17.3344 |
| 2009481817589551872 | 54.0687 | 23h13m21.0219s | +57d11m03.5485s | 14.8521 |
| 2009481748875803776 | 31.6585 | 23h13m20.403s | +57d10m31.7147s | 19.1224 |

A.15 τ Ceti

There is Hubble data for τ Ceti in MAST, though there are no images of the position in 2026. The closest in the archive is an image from 1997 in the HLA, hst_06887_35_wfpc2_total_wf_drz.fits, in which the 2026 position is just off of the image edge. The position of the target in 2000 is shown in green, while the 2026 position is beyond the image edge and marked in red.

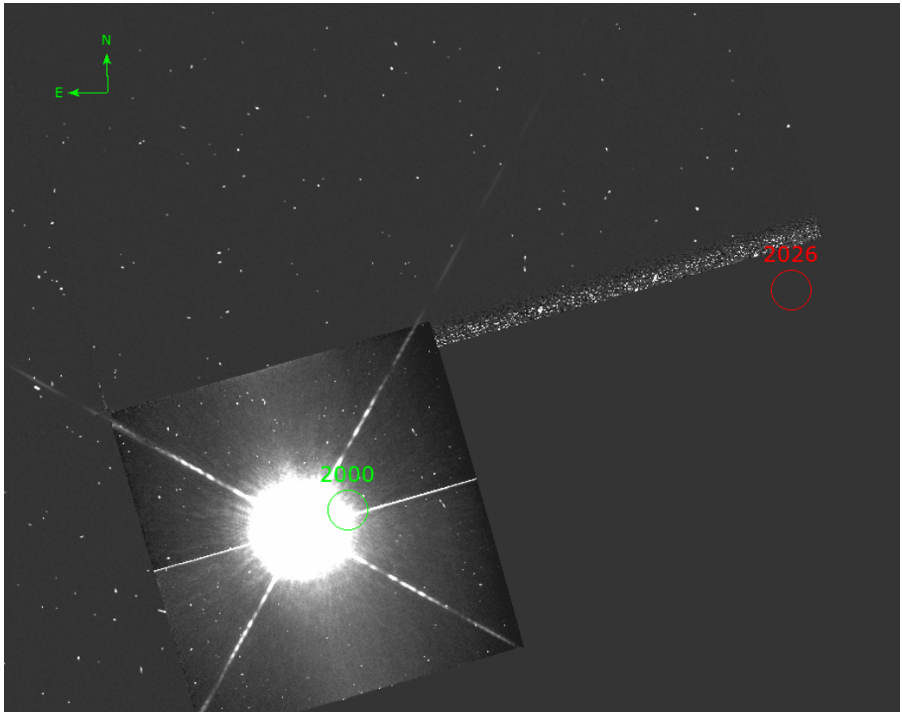

*Figure 42 Hubble WFPC2/PC HLA image of τ Ceti taken in 1997. The circle in green shows the 2000 position, while the circle in red shows where the star will be in 2026.*

There are two parallel fields with WFPC2 in the F606W filter. A visual examination of these fields in the Hubble Legacy archive shows that the fields are almost entirely galaxies or very faint point sources. Some of the sources are spurious (multiple hits on a single galaxy, for instance). Taking the field that allowed us to narrow down the field by the number of images in which the sources were identified, and ensuring that the sources were found in more than one image, there are 67 identified sources.

*Table 23 Comparison of total sources in one WFPC2 combined image (total and broken out by CI value) compared to the Gaia catalog for the area around τ Ceti.*

| Catalog being used | Area of coverage (square arcsec) | Total number of sources | Source density per square arcminute |
|---|---|---|---|
| Total of WFPC2 | 5.7 | 67 | 11.8 |
| Sources with CI < 1.2 | 5.7 | 8 | 1.4 |
| Sources with CI > 1.2 | 5.7 | 59 | 10.4 |
| Gaia (within r=18 arcmin) | 1017.9 | 685 | 0.67 |
| Besançon | 36000.0 | 169120 | 4.7 |

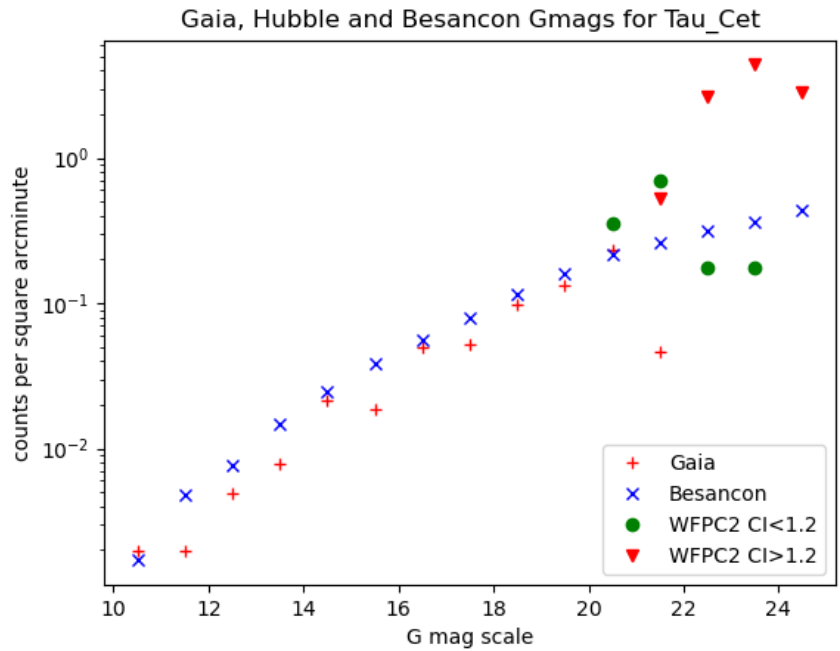

*Figure 43 Comparison of Gaia, WFPC2 and Besançon source counts per magnitudes per square arcminute near τ Ceti. The WFPC2 data is separated out into point sources (CI < 1.2) and extended sources (CI ≥ 1.2).*

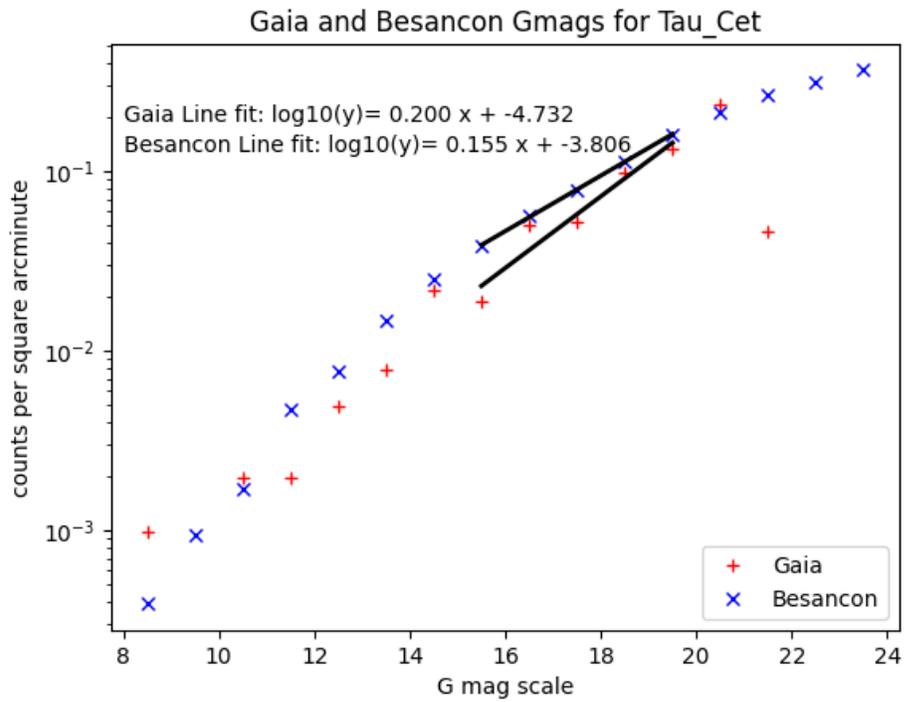

*Figure 44 Gaia counts per magnitude bin per square arcminute compared to Besançon simulations of the same for τ Ceti.*

The position of τ Ceti in 2026 is 01h44m00.9811s -15d55m52.7167s as calculated from coordinates and proper motions found in SIMBAD in J2000 coordinates.

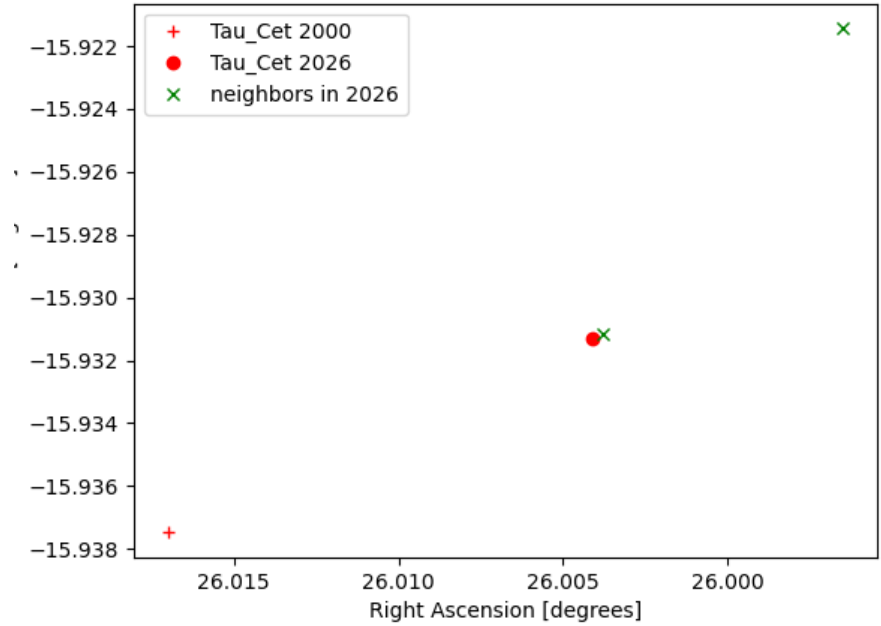

*Figure 45 Positions of τ Ceti in 2000 and 2026 compared with positions of other Gaia sources within one arcminute.*

*Table 24 Positions of stars within one arcminute of τ Ceti in 2026. Target star in bold italics.*

| Source ID | Distance (arcseconds) | RA | Dec | G mag |
|---|---|---|---|---|
| ***2452378776434276992*** | ***1.07258*** | ***01h44m00.9136s*** | ***-15d55m52.2649s*** | ***3.1336*** |
| 2452379051311170304 | 44.2041 | 01h43m59.156s | -15d55m17.2063s | 17.7728 |